碩士學位論文

# Convolutional Neural Network-based Optical Camera Communication System for Internet of Vehicles

차량 인터넷을 위한 컨볼루셔널 신경망 기반
광카메라통신 시스템

**Graduate School, Kookmin University**

**Department of Electronics Engineering**

**Amirul Islam**

**2016**

# Convolutional Neural Network-based Optical Camera Communication System for Internet of Vehicles

*by*

*Amirul Islam*

A thesis Submitted to the Department of Electronics Engineering, Graduate School, Kookmin University in partial fulfillment of the requirements for the degree of Master of Science

*Supervised by*

## Professor Yeong Min Jang

2017 년 7 월

## Graduate School, Kookmin University
## Department of Electronics Engineering
## 2016

# Convolutional Neural Network-based Optical Camera Communication System for Internet of Vehicles

A thesis Submitted in partial fulfillment of the requirements for the degree of Master of Science

*by*

*Amirul Islam*

July 2017

This is certified that it is fully adequate in scope and quality as a thesis for the degree of Master of Science

Approved by

---

Professor Sang-Chul Kim (Chair, Thesis committee)

---

Professor Sunwoong Choi (Member, Thesis committee)

---

Professor Yeong Min Jang (Thesis supervisor & member, Thesis committee)

## Graduate School, Kookmin University
## Department of Electronics Engineering
## 2016

# Amirul Islam 의 碩士學位 請求論文을 認准함

## 2017 年 7 月

審查委員長　　　金 尙 澈 ㊞

審查委員　　　　崔 善 雄 ㊞

審查委員　　　　張 暎 民 ㊞

國民大學校 一般大學院

*Dedicated*

*to*

*My Family*

# Acknowledgment

The thesis is a production of some of my beloved persons. At the very beginning, I want to explicit my obligation to my graduate supervisor, Professor Yeong Min Jang, for his immense guidance during my career in Kookmin University. Without his enormous backing, it might be difficult to commence my study. I am very grateful to have him as my graduate supervisor. Along with this, I prefer to acknowledge the contribution of all the course teachers who help to learn new concepts and technologies. My heartfelt gratefulness belongs to the committee members of thesis Professor Sang-Chul Kim, and Professor Sunwoong Choi who dedicate their precious time and determination to assess my dissertation. Their reviews and observations have enriched the contents of my research in a great margin.

Kookmin University deserves my gratefulness for giving me the chance to perform study. Special thanks to Department of Electronics Engineering for the continuous support during my study. Besides, it is my great privilege to have a great encouraging environments provided by the Government of the Republic of Korea who demand my gratefulness.

However, I am fortunate to be a graduate researcher of Wireless Networks & Communication Laboratory (WNCL) where I have developed my research skills and explored my ideas. I am so thankful to all of my lab member, either present or former, especially Md. Tanvir Hossan, Mohammad Arif Hossain, Trang Nguyen, Chang Hyun Hong, Nam Tuan Le, Eunbi Shin for their great supports all the time. My generous acknowledgement to Dr. Mostafa Zaman Chowdhury for his continuous support and direction. I want to provide my acknowledgement to all of my Bangladeshi friends living in Korea, for their boundless care and support during my stay in Korea.

Finally, I am indebted to my beloved parents and my only younger brother who are supporting and encouraging me all the time.

# Table of Contents

# List of Figures

# List of Tables

# Acronyms

| Abbreviation | Full Form |
|---|---|
| APD | Avalanche Photodiode |
| AV | Automotive Vehicle |
| AWGN | Additive White Gaussian Noise |
| BER | Bit Error Rate |
| CA | Collision Avoidance |
| CNN | Convolutional Neural Network |
| D2D | Device-to-Device |
| DoG | Difference-of-Gaussian |
| FOV | Field of View |
| fps | Frame Per Second |
| FSK | Frequency Shift Keying |
| GPS | Global Positioning System |
| TG7m | IEEE 802.18.7m Task Group |
| ICP | Iterative Closest Point |
| ICT | Information and Communication |
| IM/DD | Intensity-Modulation or Direct-Detection |
| IoT | Internet of Things |
| IoV | Internet of Vehicles |
| IS | Image Sensor |
| ISC | Image Sensor Communication |
| ITS | Intelligent Traffic System |
| LBS | Location Based Services |
| LED | Light Emitting Diode |
| LiDAR | Light Detection and Ranging |
| LoS | Line of Sight |
| LTE | Long-Term Evolution |
| M2M | Machine-to-Machine |
| MIMO | Multiple-Input Multiple-Output |
| NCC | Normalized Cross Correlation |
| NFV | Network Function Virtualization |
| NLOS | non-LOS |
| NN | Neural Network |
| NOS | Network Operating System |
| LOS | Line-of-Sight |
| OCC | Optical Camera Communication |
| OFDM | Orthogonal Frequency Division Multiplexing |
| ONF | Open Networking Foundation |
| OOK | On-Off Keying |
| OWC | Optical Wireless Communication |
| PD | Photo Diode |
| PLC | Power Line Carrier |
| S2-PSK | Spatial-2-Phase-Shift Keying |
| SIFT | Scale-invariant Feature Transform |
| SNR | Signal to Noise Ratio |
| QoS | Quality of Service |
| RADAR | Radio Detection and Ranging |
| *ReLU* | Rectified Linear Unit |
| RF | Radio Frequency |
| ROI | Region of Interest |
| SAD | Sum of Absolute Differences |
| SDN | Software-defined Networking |
| SFD | Start Frame Delimiter |
| SSD | Sum of Squared Differences |
| UFSOOK | Undersampled Frequency Shift On-Off Keying |
| V2C | Vehicle-to-Cloud |
| V2I | Vehicle-to-Infrastructure |
| V2P | Vehicle-to-Pedestrians |
| V2R | Vehicle-to-Roadside Unit |
| V2S | Vehicle-to-Sensors |
| V2V | Vehicle-to-Vehicle |
| V2X | Vehicle-to-Everything |
| VANET | Vehicular ad-hoc Network |
| VLC | Visible Light Communication |
| VoI | Vehicles-of-Interest |

# Abstract

## Convolutional Neural Network-based Optical Camera Communication System for Internet of Vehicles

*Amirul Islam*

Department of Electronics Engineering

Graduate School, Kookmin University

Seoul, Korea

The evolution of internet of vehicles (IoV) and the growing use of mobile devices with the development of the Internet of Things, demand has grown for alternative wireless communication technologies. As a promising alternative, optical-camera communication (OCC) has emerged that uses light-emitting diode (LED) and camera as transmitter and receiver respectively. Since LEDs and cameras are exploring in traffic lights, vehicles, and public lightings, OCC has the potential to handle the transport systems intelligently. Though some technologies have been proposed or developed, these are not mature enough to uphold the huge requirements of IoV. However, most of the OCC applications are limited to single vehicle and there has been limited focus on the use of multiple vehicles detection (spatial) or fast processing (temporal) systems. Also, there has no system to challenge with the bad weather condition. In this research, a vision camera and high-speed camera has been proposed to provide multiple vehicle detection and fast data processing. Here, a convolutional neural network (CNN) based vehicular OCC system has been introduced to guarantee communication to maintain communication at the adverse condition and near-infrared is proposed in addition to visible lights to provide long range and secure communication. To support IoV communication, software-defined networking (SDN) has been proposed. Finally, the results represent the required conditions for vehicular OCC system analysis and improved performance demonstration using the proposed intelligent system.

# Chapter 1
# Introduction

## 1.1 Introduction

We are living in a technologically developed world where everything is going to be connected with each other. In recent years, the number of sensor-enabled technological devices in our daily life (e.g., smartphones, laptops, music systems, televisions, vehicles, traffic lights, and others) has increased greatly. The universal network framework for these devices constitutes the basis of a future Internet of Things (IoT), to which 25 billion "things" will one day be connected, of which a significant portion will consist of on-road vehicles. This growth will create a challenging but profitable market for future connected vehicles [1]. As more vehicles are connected to the IoT, conventional vehicular networks are mobilizing into the Internet of Vehicles (IoV). Fig. 1.1 illustrates IoT opportunities for next-generation information and communication (ICT) areas.

IoV represents the evolution of vehicular ad-hoc networks (VANETs) for providing "smart transportation" [2]. The primary purpose of VANETs was to improve efficiency and traffic safety of the vehicular network through communication that connects every vehicle by a router or mobile node which in turn, create a large network [3]. As a consequence, if any vehicle drops out of VANET network signal coverage, another vehicle can participate by creating a new network. However, the commercialization interests of VANETs have not emerged sufficiently, regardless of having massive possibility to ensure vehicular safety at less operating cost [4]. VANETs are also limited in the number of connected vehicles and mobility support, because they cover very small areas and cannot provide global services for the desired applications. However, with safety being the prime goal in the continuous modernization of vehicles and road infrastructures, growing traffic casualties throughout recent years have become a serious concern. Thus, reliable cooperative vehicular communication can introduce a new era by reducing traffic casualties [5].

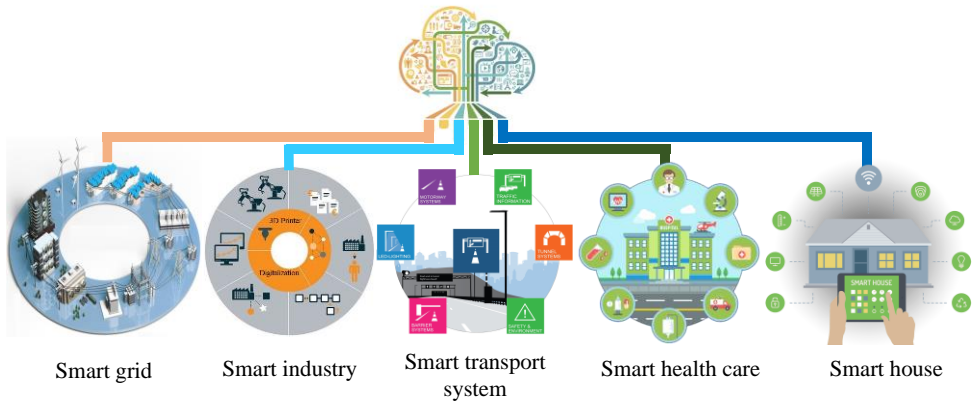

| Smart grid | Smart industry | Smart transport system | Smart health care | Smart house |

Fig. 1.1 The smart IoT for ICT convergence

To avoid collisions and accidents with vehicles, information from the surrounding vehicles and infrastructure is necessary for highly accurate and precise localization, as well as maintaining communications. The concept of establishing communication between them is promising concepts and communication between automotive vehicles (AV) is taking tremendous attention from both academia and industries, with many researchers attempting to develop the best automotive solutions. However, due to the absence of commercially available vehicles, with most of using fixed and standardized advanced technologies, it is unclear which technology is optimal. In addition, new automotive technologies must be implemented considering the regulations on the use of autonomous vehicular technology in order to ensure human safety and user trust on the machine. Google's AV has driven for more than 500,000 miles without a crash. However, the system relies on gathering information on driving conditions from around the world and on increasingly complex and sophisticated algorithms to process the sensor data and control the vehicle. Correspondingly, it requires a high amount of computational power to run in real time. Certain AV systems use combinations of sensors; however, these can mutually interfere when collecting sensor data.

The first technologies developed for sensing objects were radar and ultrasound. These two techniques were first used outside of cars to provide automated parking. Though, these technologies could be applied for detection of vehicles or pedestrians.

Moreover, the global positioning system (GPS) has a long positioning range [6] but cannot provide orientation information about the surrounding vehicles or infrastructures. Using radio waves in GPS-based positioning systems creates some problems such as radio interference error, reduced security, and multipath propagation. Its accuracy can also be hampered by the blockage of signals by tunnels, urban canyons, or dense trees. Moreover, GPS does not provide a reliable vehicular positioning because classification of the vehicle locations on various lanes are not satisfying enough. In addition to GPS, other localization technologies, such as light detection and ranging (LiDAR), light-emitting-diode detection and ranging (LEDDAR) which are mainly based on lasers, have been proposed for positioning or ranging applications. Generally, LiDAR uses light pulses from a laser source and works on the principle of time-of-flight (ToF) [7]. LiDAR can be used as an alternative approach for measuring distance between vehicles [8]. Unfortunately, LiDAR is harmful to humans, very expensive, and often very heavy; the cost of the LiDAR system can sometimes be more than that of the vehicle. More importantly, it does not have any mechanism to establish communication within the surrounding environments (e.g., vehicles, infrastructures). As information exchange between the surroundings will be essential for future intelligent transportation systems (ITS), its integration has become one of the toughest challenges to the development of ultra-reliable AV systems Thus, an optical-identification system is imperative for next-generation intelligent AV that can directly deliver the vehicles identification efficiently.

The optical spectrum can serve as a good resource for wide-band wireless communications. Optical wireless communications (OWC) are attractive for next-generation communication due to their applications to different emerging services. There are two main advantages to OWC: the potential large-transmission bandwidth due to the high-frequency carrier, and the communication security due to lack of radio-frequency (RF) radiation. Applications of this technology include 3D localization, kinetic camera-based ranging or distance measurements, various intelligent applications (e.g., virtual reality, augmented reality), different digital or display-signage applications [9] and, more importantly, autonomous-vehicular applications [10], [11]. OWC operates in three spectrum bands: the infrared, the visible, and the ultra-violet. Daily OWC applications, which typically use the visible-light spectrum can be achieved using LEDs as transmitters and photodiodes (PDs) or avalanche photodiodes (APDs) as receivers known as visible light communication (VLC). Much of VLC's potential lies in its use of light-emitting diodes (LEDs). Over the past few decades, the application of LEDs are accelerating in lighting infrastructures due to the advantages of energy efficiency, cost effective, and extended lifespan. Besides maintaining the essential functionality of lighting applications, LEDs can be triggered at a very high speed/rate and various levels of light intensity that allows data to be modulated through LED speed in a manner not detectable to the human eye [12], [13]. The VLC technology has received considerable attention in various fields of research by adopting the combined functionalities of both lighting and communication [14], [15]. Although most of these studies have mainly focused on indoor positioning, clarifications to difficulties, including communication resolution and implementation method have not been realized yet. VLC is mainly used in indoor environments for such applications as handheld terminals, robots, and intelligent furniture and appliances. Due to the non-coherent characteristic of LED-based VLC signals, signal processing using intensity-modulation or direct-detection (IM/DD) has been adopted.

Although radio frequency (RF)-based communication mechanisms (e.g., cellular, Wi-Fi, and sensor networks) are an essential part of existing wireless communication systems, such technologies have limitations that can be overcome through the use of LED-based VLC technology. Advantages of VLC over RF- and laser-based systems (e.g., LiDAR system) includes longer lifespans, low price and implementation cost, a license-free spectrum, and enhanced security owing to its line-of-sight (LOS) properties. More importantly, visible light does not pose any potential harm to human bodies or eyes, is not affected by electromagnetic interference (EMI), and allows for the smooth implementation of multiple input multiple output (MIMO) scenarios. Furthermore, it is easier to integrate VLC with the existing vehicular communication systems at a minimum additional cost and without any significant infrastructure changes because the application of LED lights are already existing in traffic lights or infrastructures and vehicles. The potential advantages of VLC have led to its recent inclusion in vehicle-to-everything (V2X) for simultaneous localization and communication [16], [17]. Although VLC can face challenges due to its LOS properties (i.e., communication links can be obstructed by objects or bad weather conditions, for example buildings, walls, rain, cloud, or fog), this is not a significant issue to consider here.

In a traditional optical communication system, the receiver often consists of a non-imaging device (i.e., PD). PDs are generally small devices that ensure a quick response. However, they operate over very small scales and the use of PDs in communication systems can reduce optical power, thereby limiting the transmission range. Correspondingly, it offers restriction due to the trade-off between transmission range and signal reception. However, these limitations can effectively be bottled-up with the development of new signal processing techniques and new materials. The use of cameras or image sensor (IS) in OWC, also known as optical camera communication (OCC) represents one innovation. Recently, cameras with visible light have been integrated into AVs for several applications including backup cameras, road-sign detection, roadside-LED detection, blind-spot detection, monitoring lane-departure. Cameras are also being introduced in vehicles for observing driver tiredness, adaptive cruise control, and distance measurements of vehicles, infrastructure. Accordingly, as a means of autonomous evolution, it is necessary to afford intelligence to the entire system in order to maintain smart decisions adaptively. The application of OCC technology to autonomous systems can be an attractive area for researchers and companies and will ensure an intelligent advanced-driver-assistance system (ADAS) in automotive environments.

LEDs are used in vehicular OCC systems to broadcast inter-vehicle information or safety information to the nearby vehicles or roadside units, while ISs are used to receive LED-transmitted information and various algorithms are used to decode information relayed by the LEDs. ISs receive the transmitted data from vehicle backlights or headlights [16], traffic lights, and signage. The important feature of an IS is that it can capture an entire scene and receive all signals within a field of view (FOV). ISs can spatially separate objects and easily distinguish among sources, allowing for the easy discrimination of the data light from interfering sources (e.g., streetlights, sunlight, and other background lighting). This ensures a high signal-to-noise ratio (SNR). In addition, OCC system provides reliable and consistent communication performance even if the distance increases [18].

Most ongoing research on autonomous localization and simultaneous V2X communication is based on visible light [11], [19] having a short distance localization (up to 10 m) without using a high-speed camera (which is very expensive); it also cannot provide reliable communication due to interference from various sources, such as sunlight or bad weather (fog, rain, or smoke). Near-infrared (NIR) light offers a promising solution to these problems. The signs of drowsiness in drivers can be detected using NIR by remotely monitoring bioelectrical signals (mainly heart rate), but much reliability improvement is needed before these systems can be implemented.

NIR offers advantages of low angular dependency. NIR can be easily used in day and night, is unaffected by bad weather conditions, and has long-distance transmission. As it is not visible, it will not be annoying to pedestrians or other drivers, and it can be easily integrated with existing visible-light-based vehicular infrastructure. NIR offers a long range and an improved detection ability than visible light, particularly in urban regions where cities are decorated with many lights. Another application of NIR systems is the adaptive control of beams, so as to automatically switch from high to less beam depending on the presence of vehicles while maintaining minimum beam forming. In NIR system, NIR-filtered camera is used to detect the presence of vehicles and decide the region-of-interest (RoI). In this research, NIR spectrum has been proposed instead of visible light considering the advantages of NIR over other spectra. Thus, the system can provide a cost effective system with simply using NIR LEDs and normal or high-speed IS which will be a very simple system.

## 1.2 Motivations

The motivations that influenced to introduce this work has been driven by several aspects. The prominent issues are specified below.

1. Continuous traffic causalities
2. Lack of established system for vehicular environment
3. Although studies on the capabilities and potentials of vehicular OCC systems are already being conducted, only a few of these have assessed the actual implementation.
4. Detection of multiple vehicles and providing high-speed communication simultaneously is necessary in the automotive environments.
5. Consideration of bad weather condition (fog, rain, snow, cloud, and etc.)
6. Maintain communication at long distance using central cloud server
7. Ensure profitable and emerging market in the coming years

## 1.3 Service Scenarios and Applications

The emerging of vehicular OCC will create lot of services and applications in the coming years. It will provide a great revolution in the AV industries. Since the IoT is starting to play an important role in everyday life, it is expected to become a major part of next-generation IoV architectures. The architecture of the IoT incorporates smart industries, smart homes, e-health, and smart grids connected through the Internet; essentially, it is anything that can be made available anytime and anywhere. In future, it will be possible to monitor the movements of trains and ships through a core network. Cellular and radio communications will also be included in these centralized networks. Fig. 1.2 illustrates multi-layer architectures of OCC networks containing both existing and upcoming technologies, such as OCC, V2X communication, device-to-device (D2D) or cellular communication, IoT clouds, and cloud-based communication. As shown in Fig. 1.2, a core network monitors the entire network and small-cell or cloud-based networks work under the supervision of this core network; in essence, the core network acts as the backbone of the entire structure. Finally, the core networks can be controlled by software-defined networking (SDN)-based cloud networks which will provide centralized services at both the user end. Because SDN provides scalable and secure centralized control of all systems through its interface.

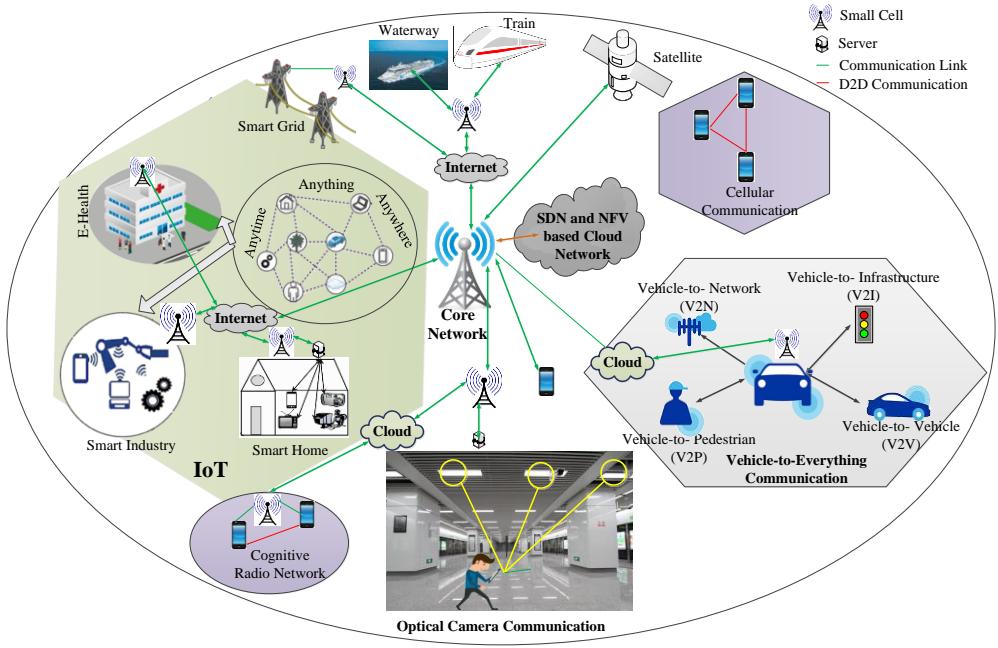

Fig. 1.2 A multi-layer architecture of future networks with OCC and IoT

OCC can also be used in tunnel environment as inside tunnel there has been a lot of accidents for lack of secure and precise localization and communications between users or workers and central controller. Fig. 1.3 shows an example of positioning and communication mechanism in tunnel environments. As shown in Fig. 1.3, the system is comprised of visible lights (e.g.; LEDs), cameras, LTE (long-term evolution) line, power line carrier (PLC), and a central controller to control and monitor the whole system. The central controller communicates with the system through LTE and the PLC. The center can transmit the details information of the current conditions by transmitting modulated signal through LEDs. The users or workers can receive the information about the vehicle or anything by only pointing ISs or smart devices. Some other application of OCC has been illustrated in Fig. 1.4 and Fig. 1.5.

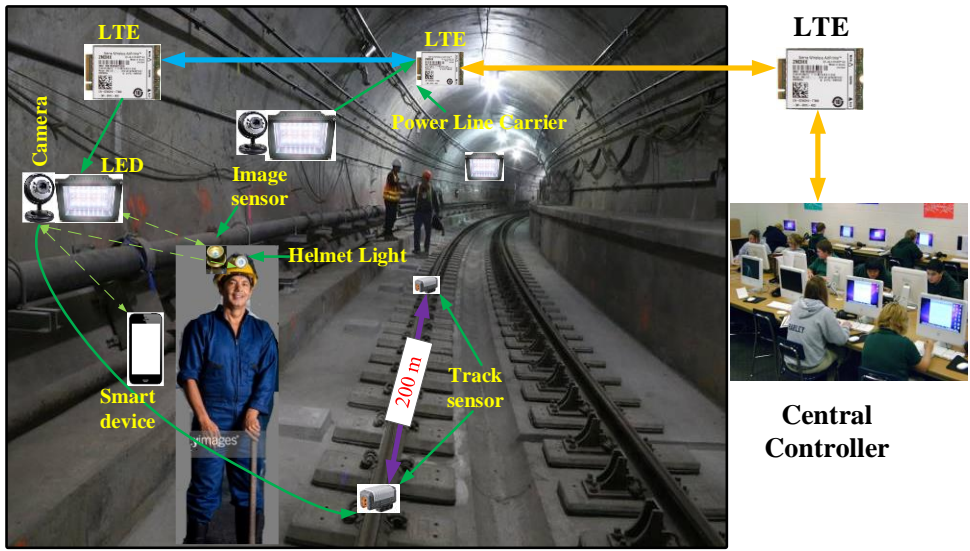

Fig. 1.3. Localization and communication in tunnel environment using OCC

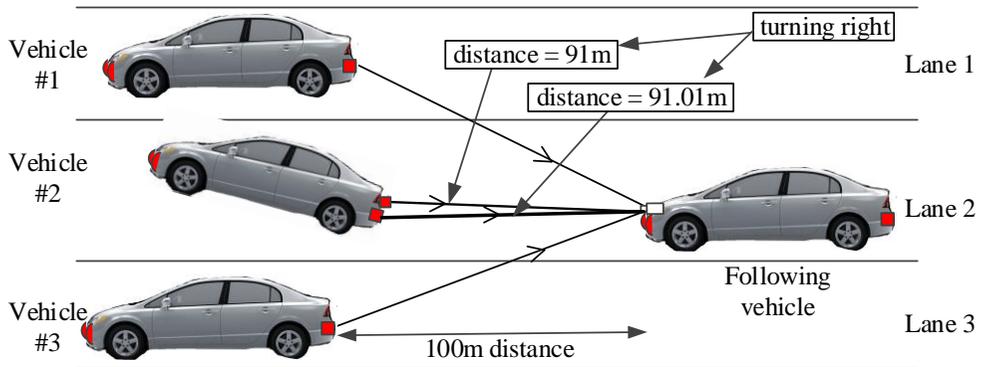

Fig. 1.4 Vehicle to vehicle communications

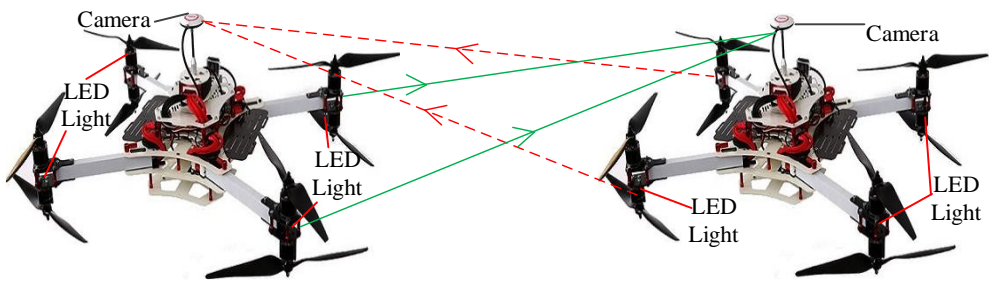

Fig. 1.5 Drone to drone communication for collision avoidance

However, intelligent optical vehicular communication can also be used for the following applications.

1. V2X communication
    i. Collision avoidance
    ii. Forward/rear collision warning
    iii. Blind-spot monitoring
    iv. Forward/rear collision warning
    v. Adaptive cruise control
    vi. Automatic emergency braking
    vii. Monitoring driver drowsiness,
    viii. Traffic sign and object recognition
    ix. Intelligent speed adaptation
2. V2X communication for safe driving and smart navigation
3. Cross traffic alert
4. Parking assistance
5. Autonomous vehicle detection and monitoring
6. Traffic jam assistance
7. Safety (reduce accidents)
8. Drone-to-drone communication for collision avoidance

## 1.4 Contributions

The work includes the development of an intelligent OCC system for AV that will ensure both vehicular-safety requirements and communication link between V2X and the central server. The prime offerings of the thesis are listed as follows:

1) Study of adaptive optical vehicular communication visible light and NIR light sources as transmitters and image sensor as receiver to provide intelligent IoV functionality.

2) Study the capability of OCC at outdoor environments which are mainly posed by the ambient noise and daylight conditions. Then, test the prospect of NIR-based IoV system to detect vehicles accurately and to satisfy vehicular-safety requirements.

3) Identify the temporal and spatial case to support multiple RoIs (multiple vehicles) detection and fast data decoding and fast processing of the decoded information of the targeted vehicles.

4) An in-depth performance analysis of OCC platform for ITS to ensure long range communication, interference avoidance, scalability of the system and least communication delay.

## 1.5 Organization of the Thesis

The remaining parts of the thesis are structured as follows.

**Chapter 2** presents the relevant works of optical vehicular communication and IoV.

**Chapter 3** describes the overview of the vehicular OCC system with the details of the transmitter, receiver, and channel model.

**Chapter 4** presents the proposed architecture based on the concept of an adaptive spatial and temporal resolution scheme with mathematical analysis and performance evaluation.

**Chapter 5** we report the proposed NIR based IoV system with detailed algorithms and performance analyses.

**Chapter 6** concludes the research along with future research directions.

# Chapter 2
# Related Works

## 2.1 Introduction

Researchers are currently striving to develop reliable systems for AVs. To make vehicles fully automotive, vehicular communication systems for smart transportation are being introduced in a process that is abetted by a continuous stream of significant technological advancements. However, consumer expectations regarding autonomous cars making their journeys easier and more comfortable will lead to increasingly complex scenarios for AV operation and therefore more challenges in their implementation, particularly with regard to safety and collision avoidance (CA), congestion, environmental concerns (i.e., energy use and emissions), road capacity, and congestion pricing [20]. Thus, AVs have a long road ahead of them until full implementation.

Severe road accident avoidance and traffic congestion are the two most important issues faced by AVs. Various researchers are attempting to develop CA systems for AVs; however, to date, such research has tended to focus on single-vehicle systems. A CA system implemented through autonomous interventions involving braking or steering has been described in a previous study [21]. The authors proposed a technique for the CA system performance estimation in worst case scenario while attempting to minimize execution errors, including early or unnecessary intervention, estimation error, and longitudinal or lateral prediction faults. They primarily focused on the measurement errors, nonlinear state predictions, and sensor and predication delays. Similar research results can be found in [22]. Unfortunately, current methodologies for AV cannot meet realistic system demands because researchers have neither focused on multiple vehicle detection nor on temporal or spatial cases.

In a real-time environment, there are multiple lanes and vehicles moving on the road at various speeds. Thus, multiple target cases should be considered while designing the systems for AVs. Unfortunately, little research has been conducted on CA in multiple vehicle situations; however, in [23], single and multiple targets for an AV have been considered. The authors built a U-disparity map based on the created RoI to characterize on road obstacles and presented a particle filter framework to track multiple targets based on the previous detection. Similar work has found in [24], where authors have used a hierarchical data association method to detect multiple targets on the road. Some other researchers are also working on multi-object detection in a different point of view [25]. But still, the amount of research result stay bottom of the bottle. Because most of the proposed technology focus on every obstacle along with the vehicle. Therefore these technologies processing some false data and the make the overall autonomous system variable.

Another important issue is that current AV are deploying a number of sensors that generate a large amount of data, which should be aggregated and analyzed in order for the vehicle to execute a decision. Data fusion for AV has been described in a previous study [26]. However, as we are concerned with issues arising from the small amounts time needed to avoid collisions, multiple vehicle detection, and, most importantly, reducing execution time to enable fast decision making, we believe that it is necessary to replace multiple sensors with a system comprising only a few sensors, e.g., ISs.

Revolutionary advancements in OWC have made the technology a potentially invaluable tool for use in the field of AV communication. Some important research progress in V2V communication using OWC with LEDs as transmitters and a CMOS IS as a receiver was reported in previous studies [10], [27]. Based on variation in LED light intensity, a flag image was generated via communication pixels with a 10-Mbps data rate. In other research, the data rate was improved to 15 Mbps with 16.6-ms real-time LED detection [11]. In [28], the transmission performance was improved to 54 Mbps with a bit error rate (BER) $<10^{-5}$ and to 45 Mbps with a zero BER. To increase the data rate, the authors proposed an optical orthogonal frequency-division multiplexing (optical-OFDM) with a transmission data rate of 54 Mbps based on the IEEE802.11p standardization.

## 2.2 Conventional VANETs

In the modern world, the number of vehicles and vehicle-assisting infrastructures is increasing rapidly, making the transportation system more vulnerable than ever. This results in more traffic congestion, road causalities, and accidents, and less road safety. To cope with the current complex traffic system, we need a unique network to accumulate vehicular-system information and ensure an effective transportation system such as VANET [29], thus providing proficient communication on the road with the help of pre-established infrastructure. VANETs connect all vehicles and infrastructure within their coverage area through a wireless router or wireless access point (WAP). The connection between the vehicle and the network can be lost when a vehicle moves away from the signal range of the network. As a consequence, a new free WAP is generated in the existing VANET for other vehicles outside of the network. Improving traffic safety and enhancing traffic efficiency by reducing time, cost and pollution are two major reasons behind the demand for VANETs.

Though creating greater opportunity in the transportation system at lower operational cost [4], VANETs suffer drawbacks such as lack of pure ad-hoc-network architecture [30], incompatibility with personal devices [31], unreliable Internet service [32], lower service accuracy, unavailability of cloud computing [33], and cooperative operational dependency of the network. Concurrently, there are a limited number of access points for particular networks. Few countries (e.g., the US and Japan) have tried to implement the basic VANET architecture but not the whole system due to the lack of commercializability. This leads to demand for more reliable and market-oriented architecture for modern transportation systems [5]. IoV can be a good candidate to meet the challenges of the commercialization problems of VANETs and the growing traffic casualties. Moreover, IoV will ensure a huge profitable market ahead for "connected vehicles" [1].

## 2.3 Current Standardization Status

Visible light was first introduced in Japan named as Visible Light Communication Consortium (VLCC) [34]. To provide more worldwide application,

VLC has been standardized in recent years. As a consequences, IEEE 802.15.7 has promoted a standardization based on VLC which was completed in 2011. An amendment of 15.7 has been proposed in IEEE named as IEEE 802.15.7m OWC Task Group which is mainly based on OCC [35]. This group is now going to make final comments on draft version D2 which will be finished in May 2018. They designed different MAC and PHY format of which a special section has been proposed for vehicular communication. Recently, a new standardization group, known as IEEE 802.15 Vehicular Assistive Technology (VAT) Interest Group, has been created mainly for vehicular scenario using OWC [36]. This standard are mainly focusing on long range vehicular technologies. There have also many ongoing standardization for ITS in ISO namely ISO/TC (International Organization for Standardization/Technical committee) 204 [37].

# Chapter 3
# Image Sensor based Vehicular Communication

## 3.1 Overview

Over the past few years, there have been many advancements in camera-based applications and services, such as multimedia, security tracking, localization [38], broadcasting [9], and ITS [10], [19]. Cameras can build image pixels projected from various light sources within their FOV. Most of the relevant research has focused upon the performances of visual communication systems based on imaging or non-imaging techniques. OWC using IS, known as OCC, is a promising technology with the functionality of LOS service and LED illumination and it has considerable superiority over existing communication technologies in wireless domain, (e.g., radio waves or single-element PDs-based communication [16], [18]). In the general OCC system, LED arrays act as transmitter which are embedded on vehicle or on traffic light and camera performs as receiver (see Fig. 3.1). Fig. 3.1 illustrates the characteristics of an IS. As shown in the figure, the data transmitted from two different LED transmitters (vehicle#1 and vehicle#2 rear LED array) can easily be captured and distinguished simultaneously using the IS. Background noise sources (e.g., sunlight, ambient light, and digital signage) can be discarded by separating the pixels associated with such noise sources. In this manner, interference-free, reliable, and secure communications can be achieved using IS.

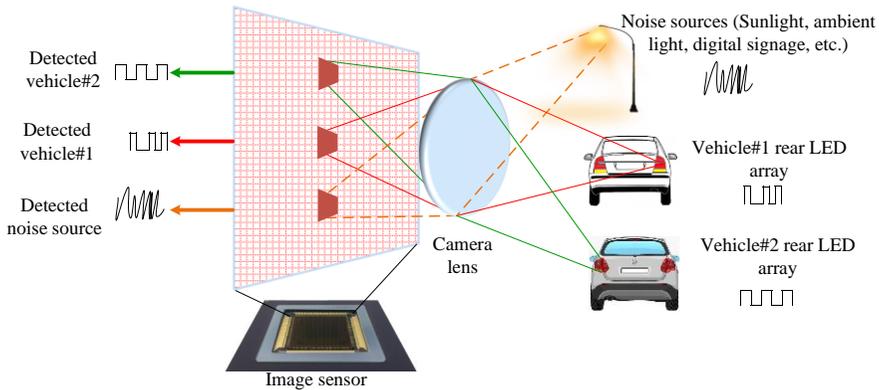

Fig. 3.1 Characteristic of an image sensor to identify vehicles or noise sources

Fig. 3.2 illustrates the overall architecture of vehicular OCC system where more than one vehicle or other noise sources serve as transmitters, while the IS serves as a receiver. In our case, the targeted or forward vehicles (transmitter) have rear LED arrays that transmit data. The LED arrays transmit vehicle safety information, including traffic information, vehicle position, LED coordinates of the targeted vehicles. Meanwhile, the IS camera receiver of the host vehicle targets the LED array and captures video frames. Then, the IS forms pixel arrays on the IS receiver focal plane through the imaging lens in order to determine the RoI within the captured images. Based on the captured images, the processor decodes information from the LED array signals using a number of demodulation techniques. The decoded information is then sent to the following vehicle's processor. Finally, based on the information broadcast from the LED arrays, the host vehicle can perform actions (e.g., reduce speed, apply braking, etc.) using a machine learning algorithm. This process is repetitively executed to improve information accuracy and obtain more data. The use of an IS considerably improves the optical energy and enables relatively high-speed, long-distance communication. Fig. 3.3 shows a simplified block diagram of an image sensor-based OWC system, which uses modulation and demodulation blocks, a communication channel, and an IS to accurately recover the transmitted data.

One of the most important advantages of the proposed technique is that a complex processing module is not required to filter out the noise sources which convey no information. These can be completely discarded using IS. More interestingly, the communication distance does not depend on the incident light power on the image pixels [18] in OCC system. In fact, distance is not affected until the image pixel size is reduced less than one pixel [39]. Thus, the communication distance can be increased to ensure stable communication. Although the optical communication distance can be further increased using zooming techniques, doing so can affect the FOV of the IS. Furthermore, in many cases, the proper lens type for particular applications has not yet been determined.

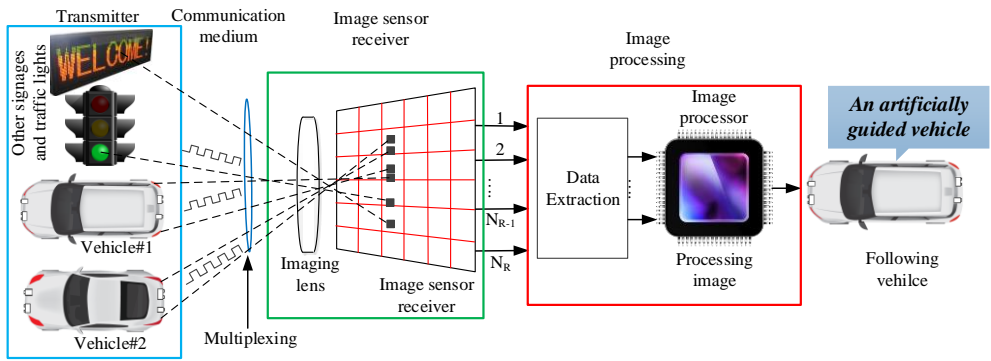

Fig. 3.2 Overall architecture of optical vehicular communication system.

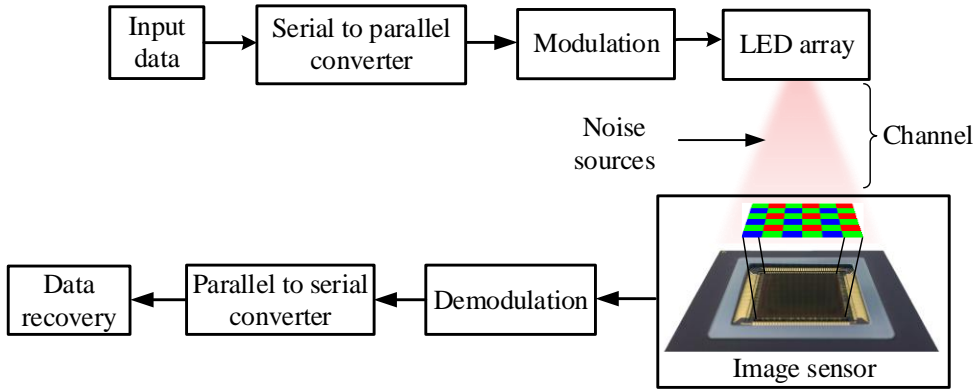

Fig. 3.3 Simplified block diagram of vehicular OCC

**Main advantages of OCC:**

1. *Interference-free communication:* IS can spatially separate the lights from various sources on its focal plane as a huge number of pixels appears in IS, which rules out the possibility of signals mixing. Therefore, communication is likely to happen even if the actual data light signals and ambient lights are present in the scene.

2. *High SNR quality*: OCC system provides very high SNR level and therefore eliminating the requirement of complex protocol for communication among multiple transmitters simultaneously.

3. *Eliminate complicated signal processing*: The unwanted LEDs can be omitted perfectly from the IS because IS can easily distinguish the communication pixels from an image, which eliminates the requirement of sophisticated signal processing.

4.  *Stable against changing communication distances:* The communication distance does not depend on the incident light power on the image pixels in OCC system. In fact, distance is not affected until the image pixel size is reduced less than one pixel.

## 3.2 NIR based OCC

Most of the recent OCC-based V2X communication systems are using visible light as a transmitter [40]. In this study, we have also proposed a vehicular OCC system using NIR spectrum. Fundamentally, NIR is employed in kinetic cameras for 3D imaging [41] as well as it is not harmful like laser light, and does not require expensive and bulky hardware to implement. NIR is less affected by interference and can be employed for long-distance communication. It can also penetrate through bad weather condition, such as snow, rain, or fog (see Fig. 3.4), which ensure long-range transmission. NIR is not visible to human eye and its signal can be detected by normal cameras. As a result, it can be used in the road environment in daytime without annoying drivers or pedestrians whilst simultaneously maintaining communication functionality. We present a comparison of NIR, visible light, and RF to prove the advantages and usefulness of NIR in Table 3.1. We propose NIR-based camera communication in vehicular environments. We transmit the emergency information of the vehicles through NIR LEDs and cameras receive the NIR signal to maintain communication between the vehicles. The transmitter and receiver, together with the channel modeling of the proposed system, is explained in the next part of this section.

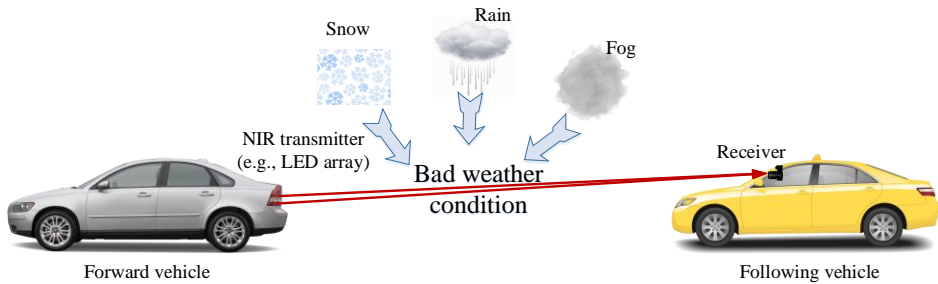

Fig. 3.4 Advantages of NIR at bad weather condition (pass through snow, rain, and fog)

Table 3.1 Comparison of NIR performance with other spectra

| Category/Properties | NIR | Visible Light | RF |
|---|---|---|---|
| Channel model | Lambertian emission, Perspective distortion | Lambertian emission, Multipath propagation | Doppler effect, Multipath propagation |
| Band / wavelength | (700-1100) nm | Unlimited, (400-750) nm | 3 Hz - 3000 GHz |
| EMI | No | No | Yes |
| Security | High due to LoS | High | Low |
| MIMO implementation | Easy | Easy | Difficult |
| Visibility range | Long distance | Short distance | No |
| Human safety | Yes | Yes | No |
| License Requirement | No | No | Yes |
| Interferences | Less | More | More |
| LOS support | Yes | Yes | No |

### 3.2.1 Transmitter

The transmitter unit is composed of an optical-NIR-emitting-diode source (typically using high-power LEDs or semiconductor laser diodes), an optical amplifier (if necessary), a modulator, optics for beam-forming, driving circuits, and a controller to control the source of data streaming as shown in Fig. 3.5. Before modulating the transmitter signals, the data from the vehicles are accumulated and channel coding is used to mitigate the receiver signal intensity fluctuations. A spatial-2-phase-shift keying (S2-PSK) [19] modulation scheme is used to modulate the NIR signal and intensified by an optical amplifier. S2-PSK modulated signals are robust and perceptible than the spatially coded signals, which are exposed to both camera-sensor resolution and partial occultation. To make the NIR safe for human eyes, we limit the average and peak-transmission powers of the optical source.

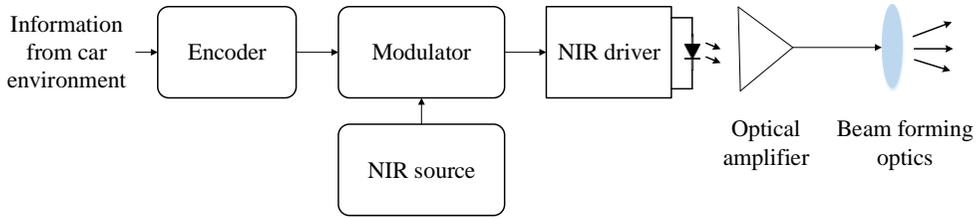

Fig. 3.5 Block diagram of the NIR transmitter

## 3.2.2 Receiver

The receiver consists of high frequency, low resolution camera equipped with a bandpass NIR filter and a decoding algorithm. The camera captures the road environments images and the decoding algorithm extracts the signal from the emitters. Fig. 3.6 illustrates the block diagram of a NIR-based OCC receiver. The whole receiver system is classified into two types: coherent and non-coherent. In the coherent system, a locally generated optical oscillator is mixed optically with the received field and then sent back to IS. On the other hand, in a non-coherent system, the emitted light intensity conveys the information. There is no requirement for a local oscillator; the IS can detect the light intensity directly. The received beam is focused onto an IS by a lens. A trans-impedance circuit converts the output current from IS into a voltage which are determined by the transmission rate, matching impedance with other receiver parts, and thermal noise generated from the receiver.

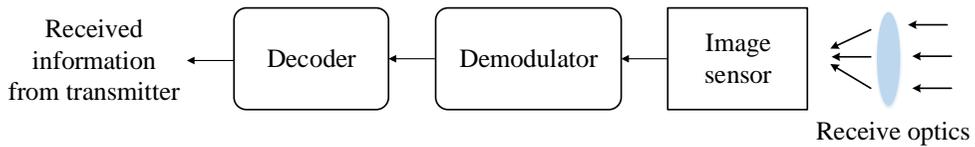

Fig. 3.6 Block diagram of the NIR receiver

## 3.2.3 Channel Model

The channel features of OCC are generally defined by the physical characteristics of transmitter and receiver. We consider additive-white Gaussian noise (AWGN) in the communication channel between transmitter and receiver. The total signal power received at the IS can be expressed by (3.1).

$$y = h_a Rx + n \tag{3.1}$$

where $h_a$ is used to explain the state of channel which represents a random variable, $R$ presents the detector responsivity, $x$ defines the intensity of the transmitted binary signal, and $n$ shows the AWGN with variance, $\sigma_n^2$. For OOK modulation, the SNR is specified by (3.2).

$$\gamma(h) = \frac{2P_t^2 R^2 h_a^2}{\sigma_n^2} = \gamma_o h_a^2 \tag{3.2}$$

where $P_t$ represents average transmitted optical power so that $x \in \{0, 2P_t\}$ have equal prior probability, and $\gamma_o = (2P_t^2 R^2)/\sigma_n^2$.

In order to measure the communication system performance, the probability distribution function (PDF) of the SNR can be derived as bellow:

$$f_\gamma(\gamma) = \frac{z^k}{2\Gamma(k)\sqrt{\gamma\gamma_o}}[\ln(\frac{1}{\sqrt{\gamma/\gamma_o}})]^{k-1}(\sqrt{\frac{\gamma}{\gamma_o}})^{z-1} \tag{3.3}$$

where $0 < \gamma \le \gamma_o$. For faded channels, the average SNR can be expressed as a function of PDF in (3.4).

$$\bar{\gamma} = \int_0^\infty \gamma f_\gamma(\gamma) d\gamma \tag{3.4}$$

## 3.3 Modulation Scheme in OCC

A range of modulation and coding schemes have been investigated for use in OCC systems because they are considered to be the key drivers in a communication system's performance. A suitable modulation scheme depends on the image sensor used, the method by which the exposure is made, and the light source used. However, constructing a classification of modulation schemes and systems solely based on the characteristics of the light source or image sensor would prevent an acceptable classification from being derived. Table 3.2 summarizes the existing OCC modulation techniques in IEEE 802.18.7m Task Group (TG7m). Some of the modulation scheme has been illustrated in Fig. 3.7 which have been explained below.

Table 3.2 A classification of OCC modulation techniques

| Modulation schemes | Flicker modulation | Flicker-free modulation | | |
|---|---|---|---|---|
| PixNet [42]<br>COBRA [43]<br>Color Shift Keying and CDMA [44]<br>Picapi camera app [45]<br>Hierarchical scheme MIMO RGB-LED [46]<br>Microsoft LightSync color code [47]<br>Structure-light-assisted OWC [48] | Nyquist sampling (NS) | | | |
| A-On-A'-Off protocol [49]<br>Space Shift Keying [50]<br>Layered Space-Time Code [51]<br>Hierarchical Rate adaption Scheme [52]<br>Erasure Coding [53] | | Frame-rate NS | | |
| TG7m UFSOOK twinkle VPPM [54]<br>UPSOOK [55]<br>TG7m S2-PSK [19] | | | RoI Under sampling | |
| OOK Manchester coding [56]<br>Light Encryption based on OOK[57]<br>Compatible OOK [58]<br>TG7m PWM/PPM code [59]<br>M-FSK [60]<br>Rolling Shutter-FSK [61]<br>Compatible M-FSK [62]<br>Hybrid M-FSK/2-PSK [62] | | | | Rolling Shutter NS |

One of the modulation schemes for OCC system i.e., Nyquist sampling method using on-off keying (OOK) modulation scheme has been proposed in [63, 43]. In this scheme sampling is performed using the formula, frame rate = 2 × pulse rate. For OOK, high brightness data bits are represented as '1' whereas low brightness bits is indicated as '0' as shown in Fig. 3.7(a). As frame rate being twice the pulse rate, two consecutive frames indicate similar data bit. One frame can indicate unclear state (i.e., not fully on or fully off) due to the LED transition. For this case, the brightness pixel represents the data bit in other frame. The maximum data rate can be between 150 bps and 1.1kbps using a high-speed camera (typically 600 fps – 4.6 kfps) and pulse rate between 300 Hz and 2.3 kHz using OOK (mono-color) modulation. Also, data rate of 5 bps was achieved using a camera of low frame rate (11 fps) named as "*Picapicamera*".

Another modulation scheme is proposed in [54], known as undersampled frequency shift OOK (UFSOOK). In this modulation scheme logic 0 (space) and logic 1 (mark) frequencies are set more than 100 Hz to mitigate flickering. These two frequencies are directed within two consecutive frames. In which space frequency represents the harmonics of frame rate of the camera having no offset frequency and mark frequency represents the harmonics of frame rate of the camera with addition of offset frequency. The selection of space and mark frequency are performed, respectively as 120 Hz and 105 Hz and sub-sampled using a typical 30 fps camera as illustrated in Fig. 3.7(b). In this modulation scheme, start frame delimiter (SFD) technique performs synchronization. The main advantage of UFSOOK is that it can be applied both in global and rolling shutter cameras though it can achieves maximum half data rate of the camera frame rate (e.g., 10 bps for a 20 fps camera). A scheme to decrypt modulated data from two OOK signal

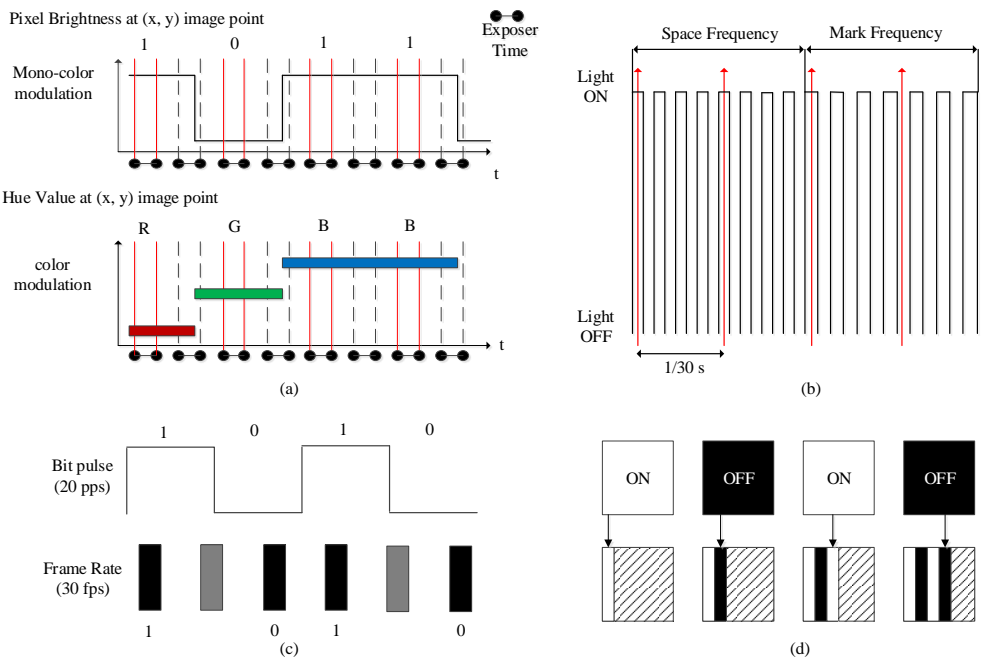

Fig. 3.7 Several OCC modulation schemes with data decoding: (a) Nyquist sampling [63], (b) USFOOK [54], (c) scheme in [64], and (d) rolling shutter scheme [56]

simultaneously was proposed in [64] where three frames were used to perform sampling and decode two data as shown in Fig. 3.7(c). It achieved maximum data rate of 20 bps using a 30 fps camera. The flickering difficulty cannot be mitigated using this process though increasing the data rate higher.

As cameras of most smart phones cameras are CMOS-based on rolling-shutter type, rolling-shutter modulation scheme is very essential. Generally, CMOS imagers are preferred compared to CCD image sensors as they guarantee least possible cost and declined size. The flickering is mitigated by LEDs switching (i.e., rapidly changing the on-off state) and varying light emitting intervals so that this flickering cannot be perceived by human eyes. The exciting frequency is made less than the scanning frequency of rolling-shutter type though keeping more than the frame rates of camera which makes LEDs lights in the state of dark ('0') and bright ('1') as exemplified in Fig. 3.7(d). Sampling based on rolling-shutter proposed for data decoding in [56]. But the decoder takes long time to process the frame and it has synchronization problems which increase the drop rate of packet [65]. Besides OOK, another scheme named as binary frequency shift keying (BFSK) is proposed to modulate signal. In this process, the LED state is represented by two frequency (i.e., $f_0$ for '0' and $f_1$ for '1'). However, in OCC system, the decoding mainly varies based on the frame rate of camera where camera frame rate depends on camera hardware design, received light intensity, API constraints, and firmware setting. Therefore, different manufacturing companies employ various frame rate for several cameras [47]. Though variable frame rate leads to bit error and poor synchronization. As a result, comprehensive investigation is essential to alleviate variable frame rate. So, in order to design modulation scheme in OCC system, the main challenges are to mitigate synchronization problem, low data rate, and flickering problem.

## 3.4 Interference Characterization of Artificial Light Sources

The most important interference source in the reception of optical signals is light reflected from multiple artificial light sources. Interference from these sources can be severe in urban areas where urban city decorated with all modern facilities, such as lots of street lights, traffic lights, and advertising boards. The artificial light sources can be categorized in three categories. These three categories of light source exhibit quite different electrical power spectra. Firstly, the light sources used for lighting purposes (e.g., decoration lights, street lights, supportive light for advertising billboards) can be a fluorescent lamp, incandescent lamp, xenon lamp, and LED lamp. These driven by AC source with a frequency of 60Hz. So, the frequency spectrum can be up to several KHz causing low-frequency interference. Secondly, light sources for static advertising purposes generally neon sign board (i.e., neon light) are driven by the ballast with the spectrum extending to tens of kHz. And the final category includes the light sources which are used for effective advertising and signaling, such as e.g. LED screens, are usually driven by sophisticated controlling circuits to display various information on the screen. This light sources create interferences in low data rate communication, have frequency spectrum of hundreds of kHz. So, the interferences can be minimized by modulating the LED light sources at very high frequency such as 0~1MHz thus improving the robustness of the system in different scenarios, or the receiver module can adaptively discard the interferences. For example, in IS based communication system IS can spatially remove the noise sources.

# Chapter 4

# Multiple RoIs Detection and Ranging

## 4.1 Introduction

In a traditional communication system, the image properties of a detected image are recovered by extracting binary values from a gray image in a simple signal or image-processing technique. As this involves the extraction of information from an entire image, the process can be more time-consuming and complex. In addition, such techniques cannot differentiate between LED light sources and other noise sources (such as sunlight, digital signage, and outdoor billboards), making it impossible to guarantee the extraction of required data effectively. In contrast, the proposed vehicular OWC scheme can easily resolve these problems using the characteristics of an IS. As explained earlier, ISs can separate communication pixels from other noise sources or interfering pixels. As mentioned in Chapter 3, the LED array transmits modulated data which can be detected the transmitting vehicle using IS receivers from other vehicles. Finally, the IS can determine the RoIs and recover broadcast information.

In this research, an adaptive vehicular communication system is introduced. In the receiver system, a vision camera and high-speed camera is combined and visible lights are used as transmitters. As shown in Fig. 4.1, the scenario of road environment can be categorized into 3 ways: Case 1: normal condition where no critical condition, Case 2: spatial condition where multiple vehicles are present and Case 3: temporal condition where one vehicle is in critical condition. The details of the scenario will be discussed in the next section. The vision camera is used for multiple vehicle detection in a spatial scenario, while the high-speed camera is used for communication purposes (i.e., decoding the information, for example, safety information, speed, information about any accident, the current movement of the vehicle, etc.). In [40], only a single high-speed camera were proposed for both temporal and spatial scenario. However, in this research, the main purpose of high-speed camera is to decode information of the targeted vehicle at very high rate. The function of vision camera is to detect multiple vehicles accurately with their region of interests (RoI), for example, LED arrays. For detection of multiple RoIs, an algorithm is suggested which combine iterative closest point (ICP) and scale-invariant feature transform (SIFT) algorithms. The SIFT algorithm detects the image features using the vision camera image, and ICP will reduce the errors obtained from the features map using SIFT algorithm. After detection of RoI, the relative distances between the vehicles are measured using vision-based distance measurement technology. From this distance information, the system can easily understand the temporal condition (i.e., which vehicle, the high-speed camera have to focus to avoid the collision and receive the broadcasting information) of the scene. To provide accurate communication in automotive applications, fast detection and processing is required. The proposed image sensor based vehicular communication system can fulfill the requirement of multiple vehicles detection and fast processing of information to support adaptive spatial and temporal condition.

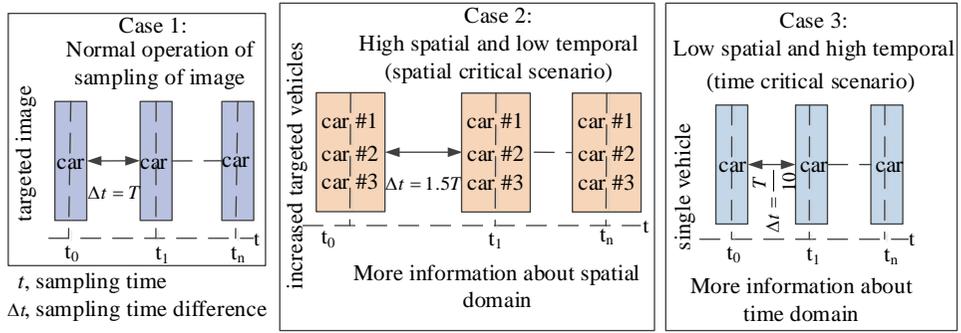

Fig. 4.1 Real road scenario in vehicular communication

## 4.2 Proposed Architecture and Modelling

Fig. 4.2 illustrates the proposed system architecture; from the figure it is apparent that the system comprises two interconnected modules: a transmission module that transmits vehicle-related information (e.g., safety and traffic information) and a receiver module that determines the RoIs of the vehicles of interest (VoI) and decodes information transmitted from the transmitter module.

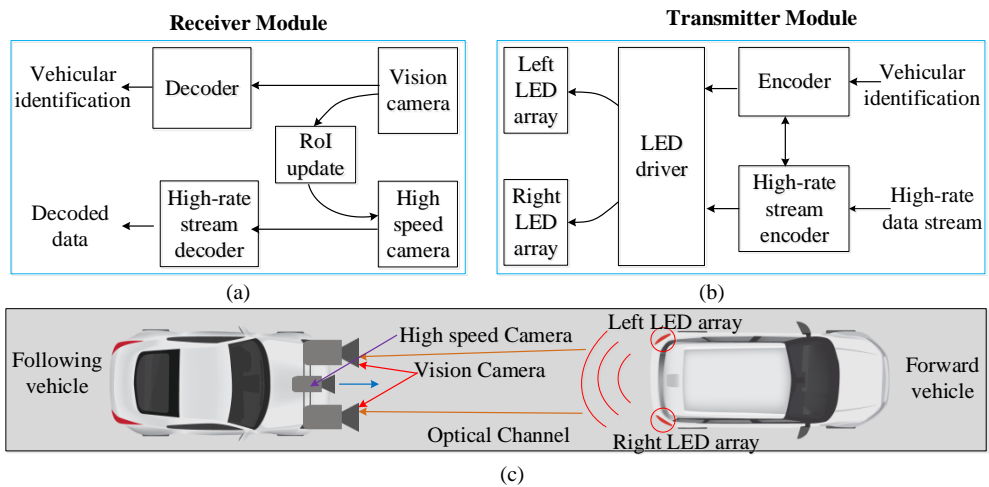

Fig. 4.2 Image sensor-based vehicular system: (a) receiver module, (b) transmitter module, and (c) overall system architecture

The transmitter module includes an encoder and an LED array unit. The encoder collects various data for sending and encodes the data into packets for relaying them to the LED array unit. The LED array unit comprises an LED array (i.e., left LED arrays and right LED arrays) and LED drivers. In this system, the LEDs can be modulated at considerably high rates that are not visible to the human eye.

Consequently, the receiver module comprises a camera receiver unit and a decoder. The camera receiver unit contains two components: (i) a vision camera receiver and (ii) a high-frame-rate camera receiver. The vision camera (stereo camera) simultaneously detects multiple vehicles and identifies the respective RoI of each vehicle (i.e., the backlight LED array of the vehicle). After vehicle RoI identification, the inter-vehicular distance is calculated using vision technology that will be comprehensively explained in the next part of this section. The high-frame-rate (up to 1,000 fps) camera is used to receive the signal from the LED array. The vision and high-speed cameras should operate in a synchronized manner, with the vision camera sending updated and accurate RoI information to the high-speed camera along with distance information. Based on these data, the proposed algorithm can adapt to recent conditions of the forward vehicles and enable the high-speed camera to focus on VoI to decode the required information.

To better understand the envisioned scenarios, the image sensor-based vehicular OCC system's functionality has been categorized into two cases: a spatial case, also known as multiple vehicles or RoI detection, and a temporal case or fast processing phase. In this research, the temporal case is identified as a critical condition. However, in the current spatial case, a scenario is considered wherein multiple vehicles are present. In this case, it is necessary to detect the RoIs of multiple vehicles in order to gather information from each vehicle and accurate detection of multiple LEDs and RoIs is imperative. For the temporal case, information about the forward vehicles (i.e., VoI) is necessary to avoid collision within a short period of time. In this phase, the image is sampled faster than in the spatial case. This fast processing of a single vehicle's information is performed according to the information of spatial detection. The proposed scheme will decide which the closest vehicle near to the following vehicle based on distance information obtained by processing multiple vehicles.

### 4.2.1 Spatial Case (Detection of Vehicles and Distance Measurement)

The primary state is the spatial case wherein multiple vehicles are present. In this case, it is necessary to process and detect multiple vehicles within single image, where the precise detection is important not the processing or sampling time. So, the processing time will be greater than the standard condition; in this study, the processing time is considered to be 1.5 times than the regular time (i.e., $\Delta t = 1.5T$), where $\Delta t$ represents the difference of sampling time between two consecutive frames and $T$ is the normal sampling time. In this section, the procedure of detecting multiple RoIs (i.e., LED array) and measuring the relative distance between a host and forward vehicle will be described using a vision camera. Vehicle detection has been assessed in many recent studies using detection features such as shadows [66] and edge histograms [67]. However, shadows are very sensitive to illumination changes and the background can influence the information badly. For this scenario, an algorithm is proposed to detect multiple vehicles accurately using SIFT and ICP. The SIFT algorithm is independent of change in illumination, noise, or orientation of image. It offers better feature points matching algorithm by extracting interest points from intensity image pixels and comparing them with its surrounding points. An error analysis model is constructed by combining the extracted point information with the ICP algorithm. The combination of SIFT and ICP algorithm (SIFT-ICP) minimizes the resulting errors in a noticeable edge.

### 4.2.1.1. SIFT Algorithm

The proposed algorithm is described details in [7]. The major steps for features extraction of multiple vehicles using SIFT are explained briefly as follows:

***Scale-Space Extrema Detection***: In this step, the interesting image feature points are categorized using difference-of-Gaussian (DoG) function. It is a close approximation to normalized Laplacian of Gaussian $\sigma^2 \nabla^2 \sigma$. DoG is calculated as,

$$D(x, y, \sigma) = (G(x, y, k\sigma) - G(x, y, \sigma)) * I(x, y) \tag{4.1}$$

In (1), $G(x, y, \sigma) = \frac{1}{2\pi\sigma} e^{-(x^2+y^2)/2\sigma^2}$ is Gaussian scale space function, *I(x, y)* is the Gaussian scale space of the input image, σ is a Gaussian kernel of variance $\sigma^2$ to provide more smooth images quickly and its domain is sampled in a particular way in order to reduce the redundancy.

(1) can be normalized into Laplacian of Gaussian as,

$$D(x, y, \sigma) = L(x, y, k\sigma) - L(x, y, \sigma) \tag{4.2}$$

Then, a DoG map is updated in the database by computing the individual DoG value for the entire image pixels.

***Keypoint Localization***: In next step, each pixel in the DoG map is plotted and compared with its eight nearest points in order to determine the local extrema of DoG function, $D(x, y, \sigma)$.

***Orientation Assignment***: The histograms of multiple orientations are computed based on the localized keypoints, where the selected histograms should be at least 80% higher than the maximum histogram. Finally, the unsearchable points are deleted to update the extrema orientation points.

***Keypoint Descriptor***: In this step, the image feature points are compared with the extrema to estimate the location and determine the best matching features.

#### 4.2.1.2. Iterative Closest Point (ICP) Algorithm

For the correction of errors, ICP algorithm adapts the extracted image feature through the SIFT algorithm. Then a model is formed by mapping the points in a data point. The model is used to reduce the sum of square errors with the closest model points and data points. The estimated errors are considered as the sum of all errors between data points and model points. Then a rotation matrix and translation vector are developed to define the new data. In ICP method, the number of the model point should be greater than the dimension of weight matrix, and the model matrix cannot be empty. More importantly, the dimension of both model and data points have to be same.

### 4.2.2 Target Achievement

After identifying the LED light sources (i.e., RoIs) on the image plane, the risk factor of action taken by surrounding vehicles is determined (i.e., finding the temporal and spatial condition of a vehicle). So, the relative distance between the vehicles is necessary to compute from the captured image. According to the distance information, the system will decide the targeted vehicle to decode the necessary safety information. To obtain accurate distance measurement and accurate positioning, the camera calibration is required. There are many camera calibration methods [68], [69]. In this research, a very simple camera calibration process is explained [70]. In this approach, the data from a single camera (vision camera) is used to determine the translation vector and rotation matrix of the vision camera. A relationship between the camera coordinate system and world coordinate system for the vision camera can be stated as follows:

$$M_{C_{mi}} = R_i M_W + T_i \qquad where, \ 1 \le i \le 2 \qquad (4.3)$$

where, $M_{C_m}$ is the camera coordinate system, $M_W$ is the world coordinate system of a plane. $R_i$, and $T_i$ are rotation matrices and translation vectors for each camera respectively. The rotation matrices and the translation vectors between two cameras are calculated using the following parameters.

$$M_{C_{m1}} = \begin{cases} R_1 R_2^T M_{C_{m2}} - R_1 R_2^T T_2 + T_1 \\ RM_{C_{m2}} + T \end{cases} \qquad (4.4)$$

$$where, \ R = R_1 R_2^T, T = -R_1 R_2^T T_2 + T_1$$

After the camera calibration, the distance from the captured images is computed. Distance information can help to make decisions for the vehicle to pay attention and information extraction. However, all incidents happen in the real world can be described in three dimensional (3D) format but the camera generates only two dimensional (2D) image. Thus, an adaptive convention algorithm is required to measure the distance from a 2D image. There are various approaches to measure the depth from 2D image. The most recent trend is to use a stereo-vision camera, rather than a single camera, in a manner analogous to vision system [71]. A stereo-vision camera consists of two cameras mounted at a fixed position on a single apparatus for (i) synchronizing the focal point and (ii) adjusting the image-focal plane of both cameras. Both cameras capture the same scene but with a slightly shifted FOV, allowing formation of a stereo-image pair. Distance measurement relies on matching the pixels in the left and right image. The following algorithm is being used to complete the task.

i) Image acquisition (i.e., input image from both left and right cameras).

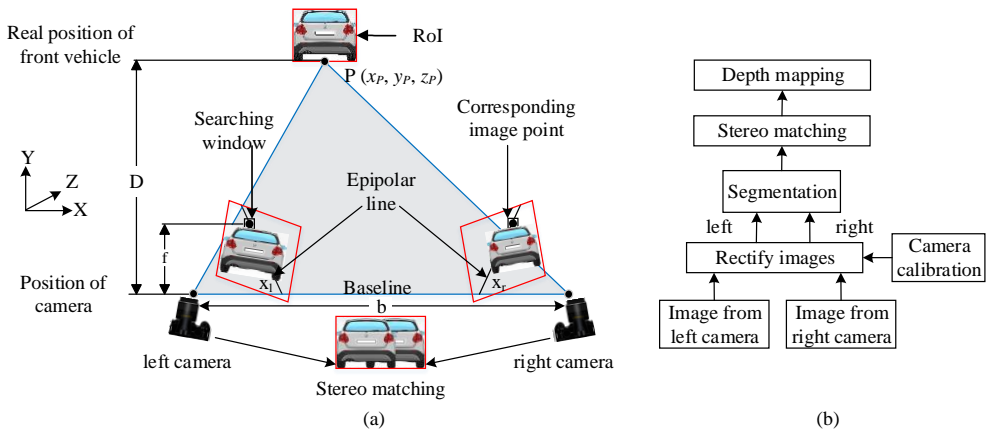

(a)

(b)

Fig. 4.3 Distance measurement: (a) using stereo images of stereo camera, (b) system platform algorithm

ii) Image rectification to align epipolar line of two camera images horizontally by using linear transformation.

iii) Segmentation for detection, recognition, and measurement of objects in images.

iv) Algorithms for stereo matching for depth calculation. There are different algorithms are being used for stereo matching, such as sum of absolute differences (SAD), correlation, normalized cross correlation (NCC), and sum of squared differences (SSD). The SAD algorithm computes the intensity differences for each center pixel $(i, j)$ in a window $W(x, y)$:

$$SAD(x, y, d) = \sum_{(i, j) \in W(x, y)}^{N} | I_L(i, j) - I_R(i - d, j) |$$ (4.5)

where $I_L$ and $I_R$ are pixel-intensity functions of the left and right image, respectively. W$(x, y)$ is a square window that surrounds the position $(x, y)$ of the pixel. The minimum difference value over the frame indicates the best matching pixel, and the position of the minimum defines the disparity of the actual pixel.

v) *Depth map estimation*: For stereo cameras with parallel optical axes (see Fig. 4.3), focal length $f$, baseline $b$, and corresponding image points $(x_l, y_l)$ and $(x_r, y_r)$, the coordinates of a 3D point $P(x_P, y_P, z_P)$ from 2D image can be determine by the following equations:

$$\frac{z_P}{f} = \frac{x_P}{x_l} = \frac{x_P - b}{x_r} = \frac{y_P}{y_l} = \frac{y_P}{y_r}$$ (4.6)

$$x_P = \frac{x_l z}{f} = b + \frac{x_r z}{f}$$ (4.7)

$$y_P = \frac{y_l z}{f} = \frac{y_r z}{f}$$ (4.8)

The depth is calculated from the disparity map using the rectified image from stereo camera. The disparity map (4.9) is determined by the difference between the x-coordinate of the projected 3D coordinate, $x_P$, onto the left camera image plane and is the x-coordinate of the projection onto the right image plane. Therefore, the disparity can be calculated from the following equation,

$$d = x_i - x_r = f\left(\frac{x_P + \frac{b}{2}}{z_P} - \frac{x_P - \frac{b}{2}}{z_P}\right) = \frac{fb}{z_P} \tag{4.9}$$

$$\text{or, } z_P = \frac{fb}{d} \tag{4.10}$$

Inter-vehicle distance can be measured as follows:

$$D = \left(\frac{f}{a}\right).\left(\frac{d}{n}\right) \tag{4.11}$$

where, $D$ is the distance between the host vehicle and the forward, $d$ is the distance between the left and right LED array units, $f$ is the lens focal length, $n$ is the distance (i.e., number of pixels) between the left and right LED array units on the image, and $a$ is the image pixel size. $D$ (obtained from the host vehicle), $f$ and $a$ are known values for any ISs. The value of n is easily obtained using simple image-processing techniques. In this manner, the system uses both the received data and captured images to provide communication and ranging functionalities.

### 4.2.3    Temporal Case (Information Decoding)

The second condition of the proposed technique is the temporal condition wherein upcoming vehicles or objects at very close distances present a potential hazard. In this case, the sampling time consideration is crucial as collisions or accidents can occur suddenly. The more information about the targeted vehicle is necessary at a given sampling time. So, the sampling time is required to increase to get more information in a very quick time. For this a high-speed camera of frame rate up to 1000 fps is proposed. So, the image can be processed quickly and the required information of the VoI can be decoded as more as possible. Fig. 4.4 presents a data decoding procedure using high speed camera in temporal condition. For temporal case, the sampling time between two frames should be faster than the spatial case in order to detect and process the VoI as fast as possible. For simplifying the simulation operation, the processing time was considered as one-tenth that of the normal time (i.e., $\Delta t = T / 10$).

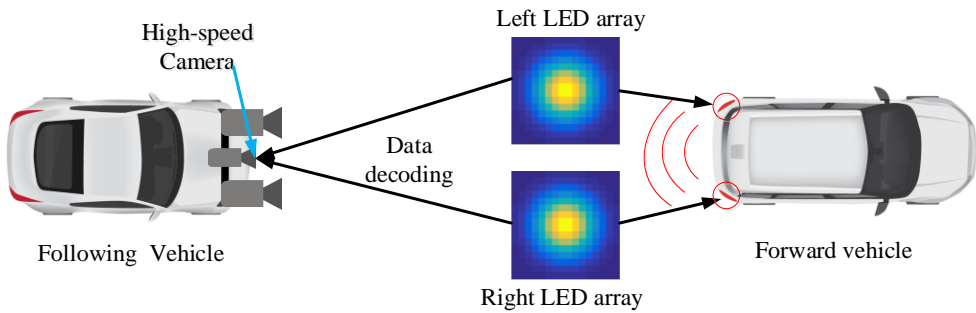

Fig. 4.4 Information decoding using high-speed camera in temporal scenario

In the proposed algorithm, the primary consideration is that the horizontal axis of the vehicle's position have to be fixed with respect to the camera position so that the camera can extract more interested image points under planer motion to develop a line of fixed points. In the next sampling, the camera will also constitute a second pair of fixed points. The two image will intersect with each other at the same point in the image plane, though the images in different sampling time can be different in nature. As the two image axes are parallel, there will be a common point at infinity where the point will intersect. As there is one fixed point among the various views, this can be used to perform the projective reconstruction on the image plane at infinity. The overall algorithm for data decoding is presented in Table 4.1.

Table 4.1 Algorithm for information decoding high-speed camera

| |
|---|
| 1) *Interest points:* extract the interest points of each image pixel. |
| 2) *Compute image homogeneity:* Estimate the homogeneity between a set of extracted feature points based on the similarity and proximity of their neighbor pixel intensity. |
| 3) *Estimation:* Repeat the step 1and 2 process until *N* samples. <br><br> i)  Select a random four consecutive samples and estimate the homogeneity *H* between them. <br><br> ii)  Measure the distance *d* from each homogeneity map. <br><br> iii) Match the uniform number of inlier with *H* by the number of correspondences so that $d < t = \sqrt{5.99}\sigma$ , where $\sigma$ is standard deviation. Finally, choose *H* with the largest number of matching inliers. |
| 4) *Optimal estimation*: Further interest point correspondences are now determined using the estimated H. |
| 5) *Iterative improvement*: Repeat the steps 3 and 4 until the two consecutive frames are stable. |

## 4.3 Performance Analysis

To validate the proposed algorithm, the proposed SIFT-ICP algorithm has been implemented in a MATLAB programming application. As a consequence, Fig. 4.5 shows the overall proposed scenario for adaptive spatial and temporal scenario wherein Fig. 4.5(a) represents the spatial case scenario with vision camera and Fig. 4.5(b) shows the temporal case scenario using high-speed (i.e., high frame rate) camera. In spatial case, three vehicles were considered where the sampling time of the targeted image is 1.5 times greater than the normal time (i.e., $\Delta t = 1.5T$ ). As discussed in the previous section, vision camera was used for multiple RoI detection and distance measurements. Firstly, the SIFT algorithm was used for feature points extraction from the intensity images. Then these extracted feature points were combined with ICP algorithm to analysis the feature points and produce an error analysis model. Finally, this combined SIFT-ICP algorithm is utilized to minimize the errors in a prominent margin.

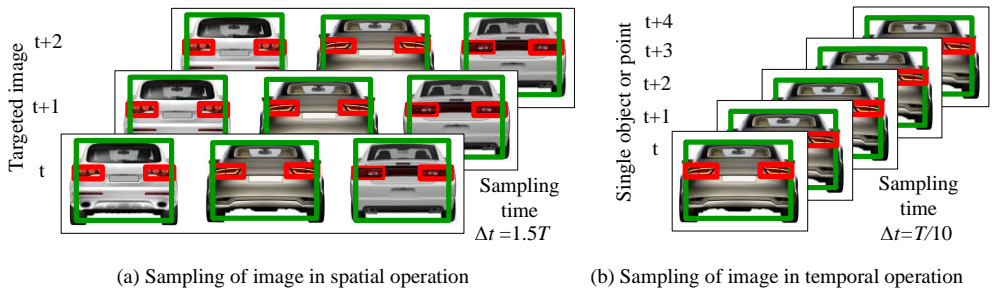

(a) Sampling of image in spatial operation       (b) Sampling of image in temporal operation

Fig. 4.5 Adaptive spatial and temporal scenario of proposed vehicular communication

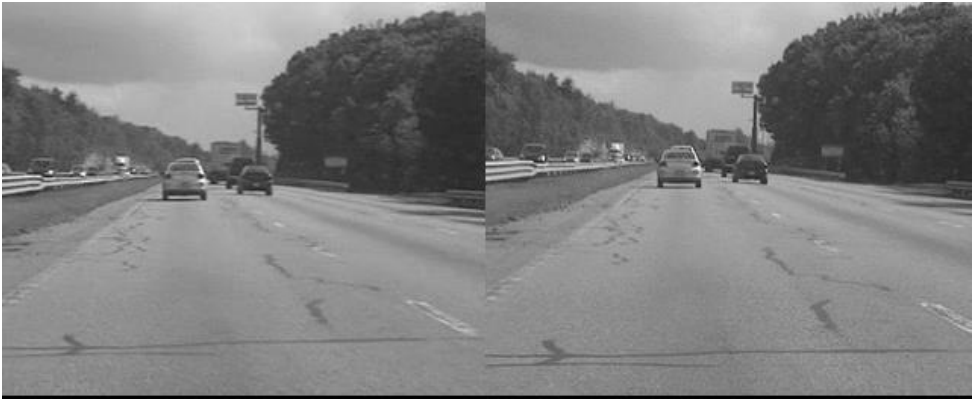

Frame 1 (left camera)                    Frame 2 (right camera)

Fig. 4.6 Two sample frames to test the proposed algorithm

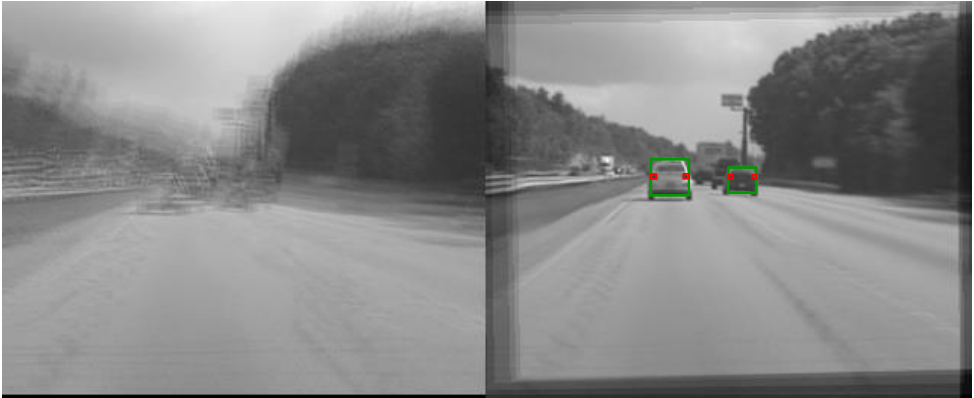

Raw mean of video frame                  Corrected errors

Fig. 4.7 Error model analysis of test images using SIFT-ICP

For RoI detection, first two frames (i.e., Frame 1 and Frame 2) of a video sequence were employed for simulation which is shown side by side in Fig. 4.6. The intensity images were used to improve the processing time and it is not important to use color image for image stabilization. A large horizontal and vertical offset are visible in the figure which generates errors. To correct the error among the consecutive image frames is the main objective of this scheme. For this, we applied *estimateGeometricTransform* function which can be done by calibrating the camera accurately. For detecting this features, SIFT-ICP algorithm was employed to a given set of point correspondences between the two frames. During computation, a given set of video frames $T_i$ (where, $i$=1, 2, 3, …., $n$) have been used. SIFT-ICP algorithm was applied to estimate the error between frames $T_i$ and $T_{i+1}$ which constitute an affine transforms, $H_i$. So, the cumulative error of a frame $i$, can be expressed as a product of all the previous inter-frame transforms which can be formulated as follows,

$$H_{cumulative,i} = H_i \prod_{j=0}^{i-1} \qquad (4.12)$$

The affine transform has six parameters but for numerical simplicity of the proposed algorithm the matrix was redefined and fitted the matrix as a simpler scale-rotation-translation transform. This new matrix (also called scale-rotation-translation transform) consists of only four parameters named as one angle, one scale factor, and two translations. So, the simplified form of the matrix can be as follows,

$$H_i = \begin{vmatrix} s*\cos\alpha & s*\sin\alpha & 0 \\ s*\sin\alpha & s*\cos\alpha & 0 \\ t_x & t_y & 1 \end{vmatrix} \qquad (4.13)$$

Then the all step applied to the full video frame to make the sequence smooth. At each step, the $H$ transform between the present frames was calculated. Fig. 4.7 show the error analysis of the video frame presented side by side (i.e., the mean of the raw video frames and of the corrected frames). The first image of Fig. 4.7 illustrates that the original image had lot of distortions in the raw mean of the video frames. The right side image represents that in the corrected frame after applying error correction method (i.e., SIFT-ICP algorithm), with almost no error. The background details has been blurred as this is not important in the vehicular condition. Finally, the results shows the efficiency of the proposed algorithm to detect the vehicles and reduce the errors for a stabilization system.

Fig. 4.8 illustrates the detailed steps for RoI detection of LED arrays. The first step acts as the convolution filter to determine the RoI in which light sources are possibly available. The second step aims to determine the accurate positions of LED light sources from the targeted RoI, as well as grouping these light sources into pair that belong to different VoIs. Each layer consists of two steps, specifically, 'convolution' and 'RoI pooling'. In convolution, also known as convolution filtering, the network attempts to label the input signal by referring to what it has learned in the past. It is advantageous for being translational invariant. Hence, the pair of light sources can shift in different positions, and the SIFT algorithm would still be able to recognize it. In another word, in RoI pooling, the RoI is down-sampled to reduce size, thereby minimizing cost. The sensitivity to noise and variations is reduced by applying RoI pooling. Noisy light sources, such as street lights without the features can be easily be discarded through this algorithm. The final RoI pooling ends when the centers of light sources are identified with the acceptable error.

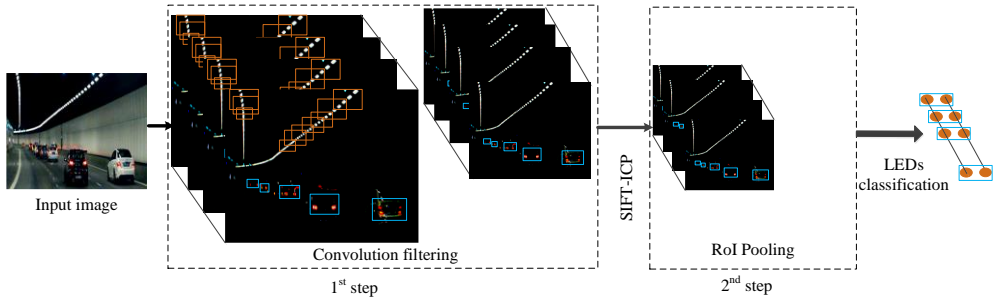

Fig. 4.8 Multiple RoIs detection and LED array classification using SIFT-ICP

For the simulation purposes, several transformation were applied, such as shift the positions of light sources to match the interest points, resize the image, rotate about the center, and mirror the original image. Fig. 4.9(a) compares the performance of LED-lights detection accuracy between general computer vision approach and the proposed SIFT-ICP approach. A computer vision-based approach processes the entire image to track LED positions, whereas an SIFT-ICP approach estimates possible positions of LEDs from previous learning and validates the correct positions from post-processing. The advantage of SIFT-ICP is that the grouping of LEDs that belong to a vehicle is simple. In addition, if LEDs are captured as OFF states, computer vision is unable to detect the exact positions of LEDs but SIFT-ICP can perform the detection accurately through estimation. The accuracy of detection depends on the distance which changes linearly since the resolution of the image is constant. A linear classification of noisy LED states has been performed to decide ON/OFF, is considerably affected by the exposure time of the camera and an additive noise; therefore, the error rate is higher than our proposed approach. Our SIFT-ICP can learn nonlinear classification to satisfy MSE of $10^{-3}$ and $10^{-10}$ within 30 and 60 iterations respectively, as shown in Fig. 4.9(b).

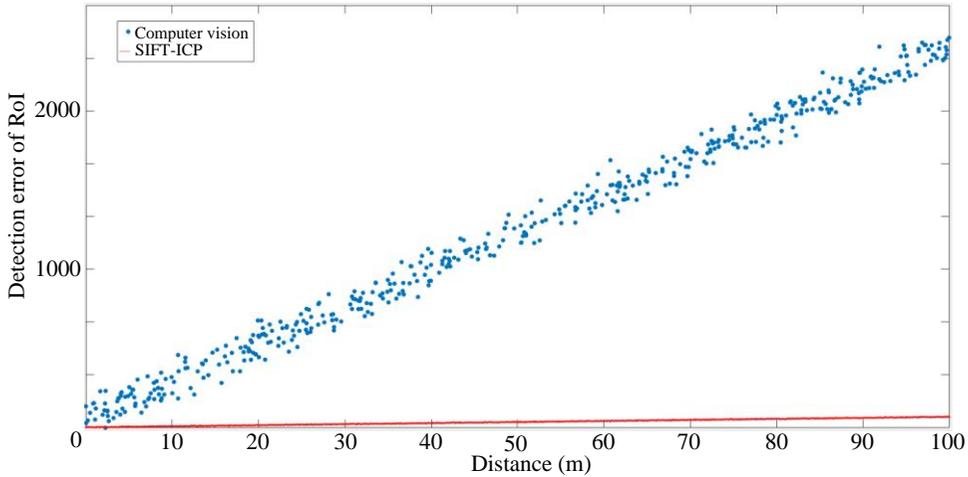

(a)

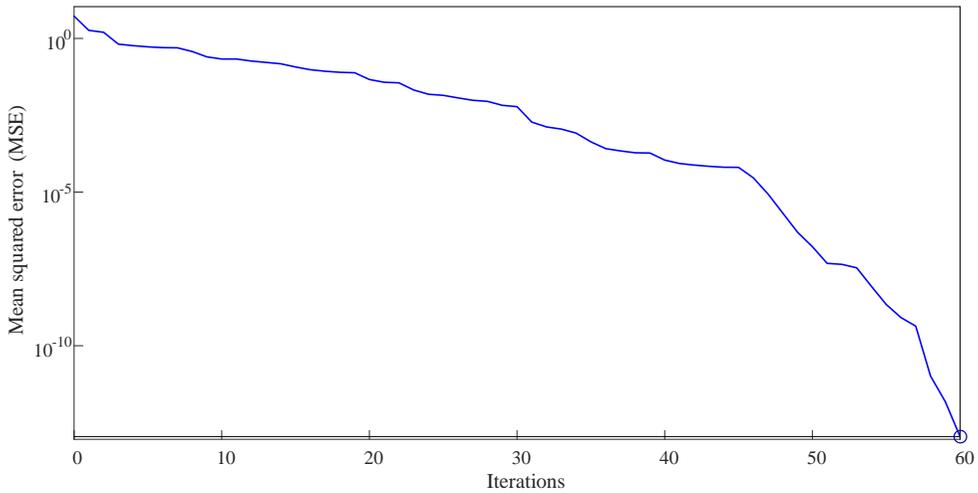

(b)

Fig. 4.9 Performance analysis of: (a) RoI detection error of SIFT-ICP algorithm in comparison with computer vision, (b) measurement errors of vehicular identification using RoI signaling.

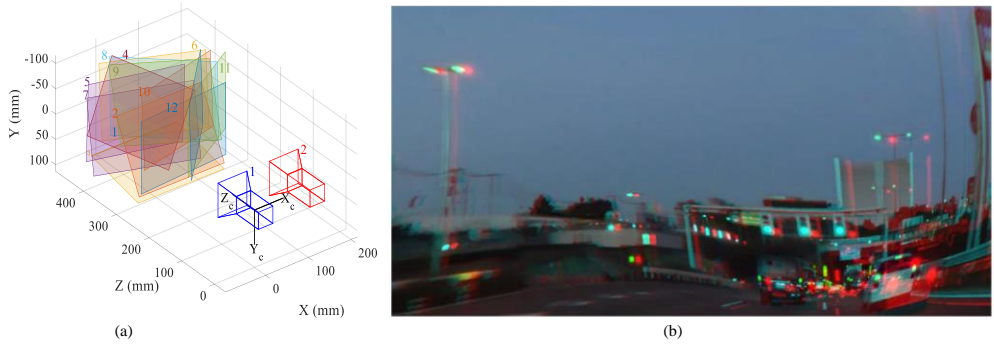

(a)                                                                (b)

Fig. 4.10 (a) Extrinsic parameters visualization of camera calibration process, (b) rectified image of the video frame in the 3D-image plane.

After detection of RoIs, the information of the vehicle is necessary to decode where two consecutive processes were followed: distance measurement and LED array pattern detection. For distance measurement, the stereo-camera approach and the MATLAB stereo-camera-calibrator app were used. Fig. 4.10(a) shows the visualization of extrinsic parameters of the camera in the 3D-image plane. In order to compute the disparity map and construct the 3D scene, the image frames from both cameras should be rectified. Fig. 4.10(b) shows the rectified image of the video frames, which are row-aligned. This will simplify the computation of the disparity map by reducing the search space for matching points to one dimension. The rectified images can be viewed using stereo red-cyan glasses, combining them into an anaglyph to obtain the 3D view.

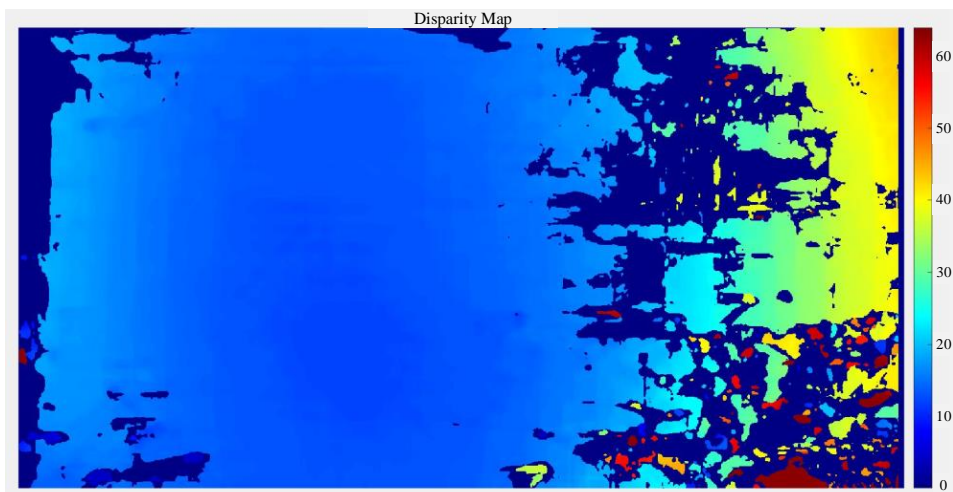

Fig. 4.11 Disparity map of the video frame in the test image

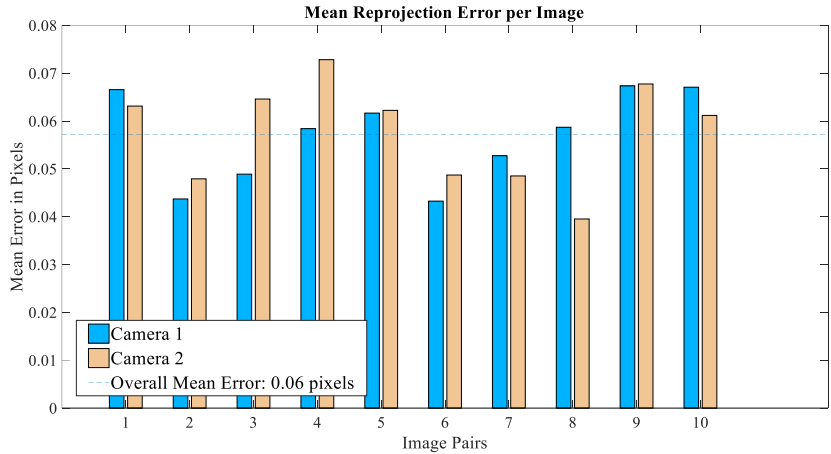

Fig. 4.12 Mean re-projection error per image pixel in image pairs

Fig. 4.11 shows the disparity map of the stereo image chosen from the video frame. From the rectified stereo images, the corresponding points located on the same pixel row are paired. The distance for each pixel in the right image to the corresponding pixel in the left image were computed. This computed distance is termed as the disparity which is proportional to the corresponding distance of the real world coordinate from the camera. Whereas Fig. 4.12 shows the mean re-projection error of the image pairs for each image pixel. For the simulation, 10 image pairs were considered. The bar graph indicates the accuracy of the calibration. Each bar shows the mean re-projection error for the corresponding calibration image. The re-projection errors are the differences between the corner points detected in the image, and the corresponding real-world points projected onto the image. From the figure, we can conclude that the average mean re-projection error is 0.06 pixels for the image pairs. The 3D-world coordinates of each corresponding pixel point were reconstructed from the disparity-map image presented in Fig. 4.13. To visualize the point cloud, the corresponding mapping pixels were converted into distance values and created a point-cloud object. Finally, a streaming point cloud viewer has been generated to present this point cloud.

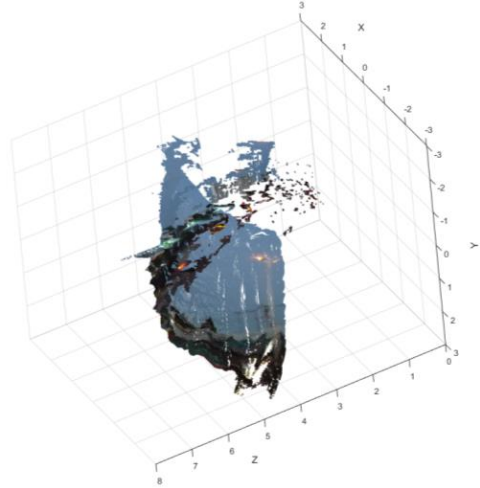

Fig. 4.13 Depth representation of the vehicular scenario in terms of cloud points

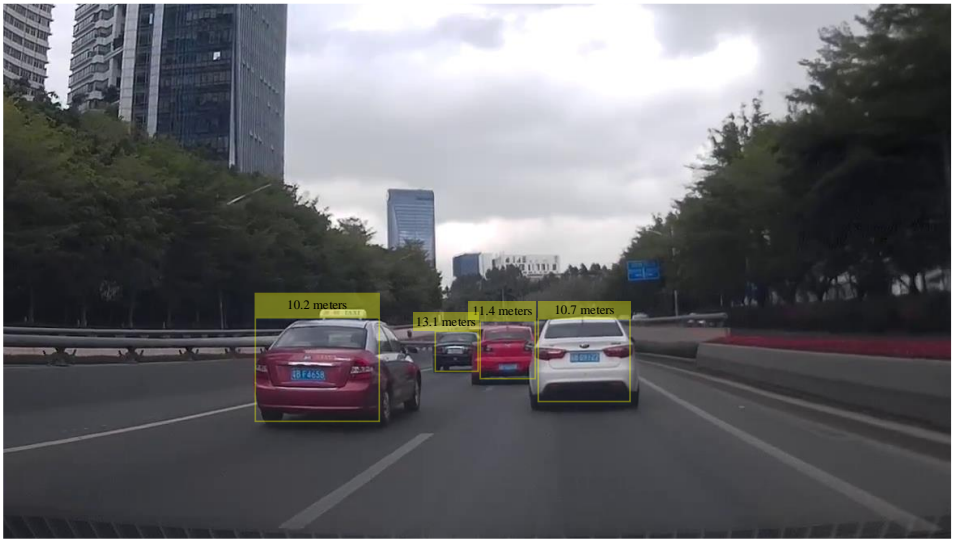

Fig. 4.14 Distance measurement using stereo camera-based depth estimation

After depth estimation, the distance of each vehicle is determined using the camera of following vehicle. The 3D world coordinates of each detected vehicle were calculated and computed their corresponding distances using (4.10). Fig. 4.14 shows the results of distance calculation. From a certain time lapse video, we found a nearest vehicle at 10.2 m distance from the following vehicle, so the system will focus on this vehicle to recognize the LED array patterns. Moreover, distance information is crucial to avoid the accidents or critical conditions.

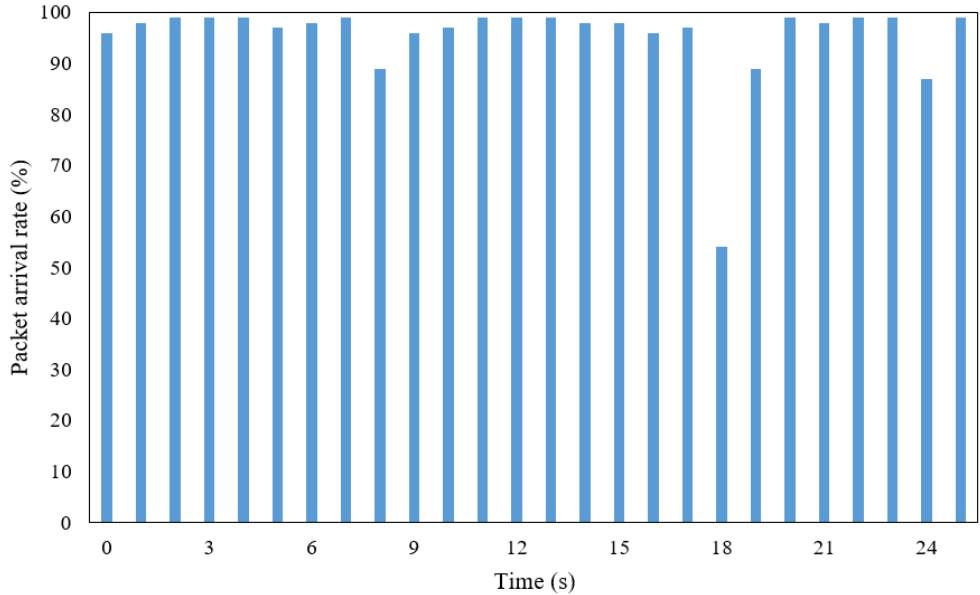

Fig. 4.15 Packet arrival rate of a video sequence using high speed camera

So, after distance measurement, the information was decoded using high-speed camera. The proposed decoding algorithm was tested in simulation. As a consequences, Fig. 4.15 shows the results of packet arrival rate in 25 seconds of a video scene using high-speed camera to decode the broadcasted safety information. The video frame was run for a time of 25 seconds to calculate the arrival rate. As shown in the figure, the fluctuation of packet arrival rate happened occasionally but the worst drop occurred at 18 seconds (54 percent). The failure of packet reception can happen due to the movement of forward or host vehicle on the rough road, or the detection error of LEDs during packet receiving or waiting. But the vehicles landed on the uneven road is one of the commonest causes of packet losses. The packet loss can be reduced by improving the LED detection rates. Though the reduction of payload can improve the reception rate, the efficiency of data transmission will reduce due to the increase in packet overhead size.

Fig. 4.16 shows relation of average throughput with varying frame arrival rate where the throughput value increased rapidly from 0 to 30 fps. Its value progressed slowly when the frame arrival rate crossed 30 fps because there had many data remaining in the queue throughout the period. The throughput became almost saturated after 90 fps. However, the transmitters can only send limited packet per frame, so the increase in frame arrival rate cannot affect the throughput after the threshold value. The average throughput is related to data transmission rate; the higher probability LEDs can send data, the higher average throughput will result. The outcome from this figure is that the adaptive method enables reliable communication even if in noisy situations when the standard system is barely feasible. Finally, the BER as a function of SNR of the image sensor-based vehicular communication system in comparison to standard system have been calculated (Fig. 4.17). As shown in the figure, the BER performance of the standard system gets worse than our proposed adaptive system. The IS has an effect on the improvement of the BER performance because it has the capability to separate the light sources (LED array) from interferences or noise sources.

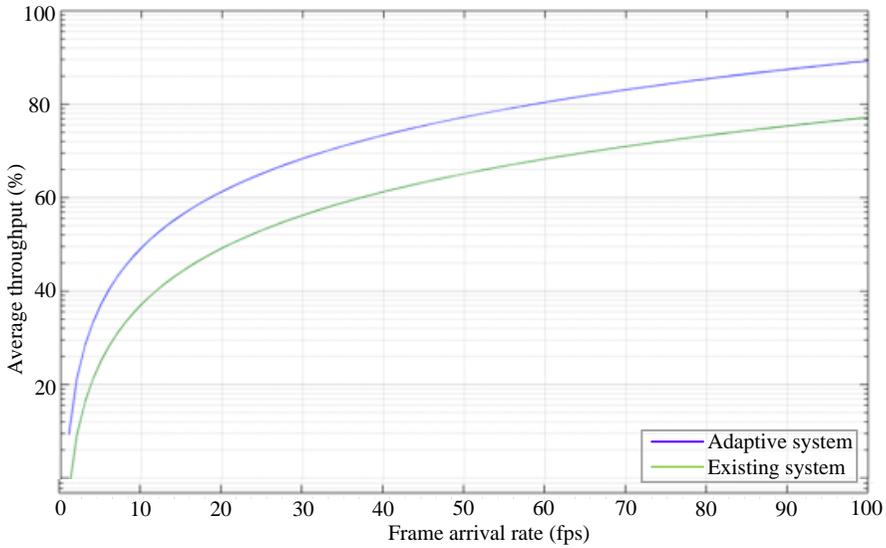

Fig. 4.16 The average throughput of the proposed adaptive system in comparison with standard system

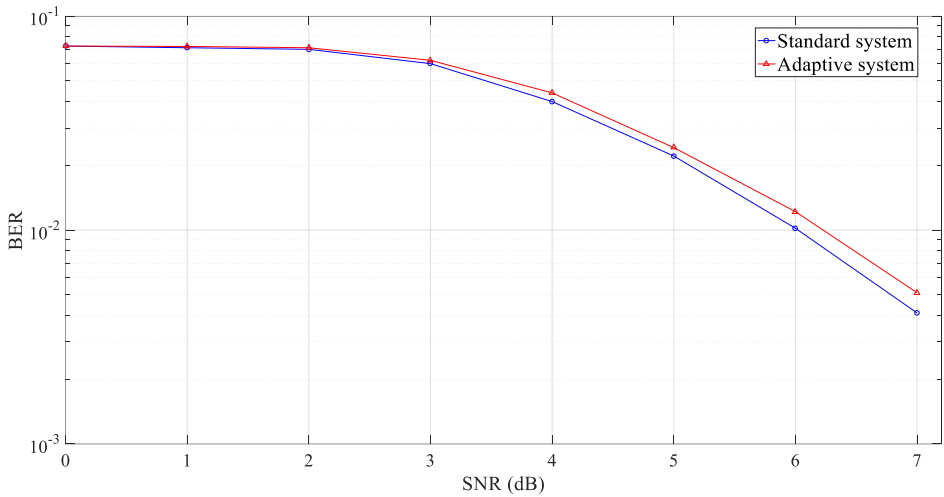

Fig. 4.17 BER vs SNR performance comparison between proposed adaptive system and standard system

## 4.4 Conclusion

In this research, a vehicular communication system for adaptive spatial and temporal conditions using image sensor-based OCC system has been introduced. Using ISs guarantee interference-free communication which is not sensitive to noise or interference. After investigating some advantages of LEDs and ISs for communication purposes, we explained the operation and functionality of an image sensor-based vehicular communication technique. We then proposed a two-phase communication method employing image sensor-based OWC communication for the spatial detection of multiple vehicles and the fast processing of targeted vehicle's data. The vision camera was used to detect multiple vehicles and measure inter-vehicular distance. Based on the obtained distance information, the transmitted information can be decoded using a high-speed camera. The decoded information using high-speed camera based algorithm has higher accuracy than other existing system. The results demonstrated simultaneous detection of three objects using a vision camera while maintaining fast processing in the temporal condition using a high-speed camera. In addition, results confirmed that the system can quickly adapt to changes in the conditions of communication environments.

# Chapter 5

# NIR-based Communication for Internet of Vehicles

## 5.1 Introduction

The IoV is an integral part of the IoT, which connects various heterogeneous networks, such as inter- and intra-vehicle networks with the Internet [2]. This heterogeneous network architecture can be categorized into five types of vehicular communication. These types include vehicle-to-vehicle (V2V), vehicle-to-infrastructure (V2I), vehicle-to-cloud (V2C), vehicle-to-sensors (V2S), and vehicle-to-personal devices (V2P) (see Fig. 5.1). It is anticipated that enhanced traffic safety, improved traffic efficiency, and implementation of supervision and control can be assured by the IoV (e.g., information exchange through inter-vehicle communications, and real-time broadcast of traffic conditions to data centers [72], [73]). Its purposes are to ensure that devices can communicate with each other and exchange information and intelligence. Moreover, the research and development of IoV technologies will integrate the automotive and information technologies together.

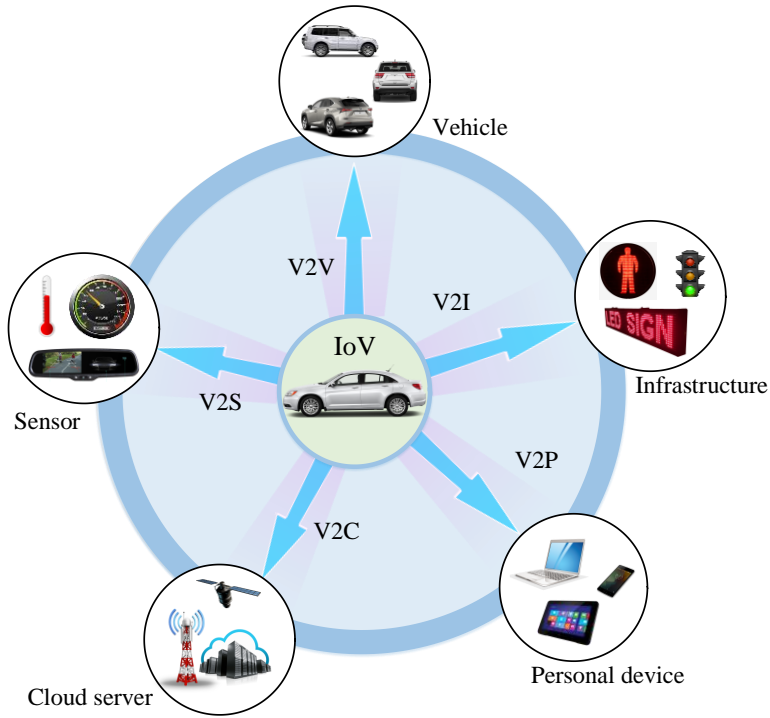

Fig. 5.1 The five types of vehicular communication in IoV.

Although many researchers and industries have proposed IoV for various purposes [74], [75], the usage of this concept is still in its initial stage. Many countries have already implemented or proposed some basic IoV services. The USA has installed security chips in vehicles to store the identify information of every vehicle in their data center [76]. The government of India has brought each and every vehicle under a network through GPS and wireless fidelity (Wi-Fi) [77]. A new next- generation transportation system named called the cooperative-intelligent transportation systems (C-ITS) has been initiated by the European Commission [78]. Korea and Australia are also interested in implementing such automotive vehicular systems in their countries [79]. Moreover, both Google and Apple are developing a smartphone-based connected-driving system. Apple has developed a system called "CarPlay," with a voice-support feature to assist the driver in using all the services of iPhone through the display of car [80]. All the aforementioned efforts are the initial steps towards the design and development of the IoV.

Lack of coordination and communication is the biggest challenge facing IoV implementation. Lack of standards also makes effective V2V communication and connection difficult and restricts use on a large scale. Thus, the challenges start with ensuring safety, simple implementation, energy-savings, and convenient and co-operative communication. Large concentrations of vehicles, lighting systems, or drones can also limit the existing RF or ad-hoc computational resources. Thus, it is of great interest to the ITS field to develop new solutions complementary to RF or other ad-hoc networks. Complementary efforts should be made to develop and enhance new platforms, which will enable analytic and semantic processing of data coming from vehicles. OCC is a candidate method for co-operative vehicular communication, enhancing driving safety, ITS, collision warning, and pedestrian detection, as well as providing range estimates for nearby vehicles. OCC is a new technology, a development of OWC that works on the same principle. OCC uses LEDs as transmitters and camera as receivers instead of PD in the case of OWC.

In section of the thesis paper, OCC system has been combined with NIR (detectable frequency bandwidth 700 -1000 nm) to provide IoV functionality, which will ensure long-range identification, remain reliable during bad weather, and not be vulnerable to partial occultation. In this case, two key elements are necessary to recognize NIR signals based on OCC technology in ITS applications: (i) the feasibility of OCC for outdoor conditions and under constraints posed mainly by the ambient noise and daylight conditions; (ii) the capability of this technology to detect vehicles accurately and to satisfy vehicular-safety requirements. Here, a NIR-based OCC system was developed by taking account of the following:

  i.   Using precise optical NIR-filtering techniques at the receiver end, allowing to increase the prototype's robustness against ambient noise.

  ii.   Using off-the-shelf components in our implementation, with the observance of the standard feasible form factor of the vehicular-lighting system.

  iii.   The performance of the NIR-based OCC system under daylight conditions and for line-of-sight (LOS) scenarios for inter-vehicle communication is evaluated using a MATLAB simulation model.

  iv.   A convolutional neural networks (CNNs) is introduced to accurately identify the LEDs' pattern and to detect this pattern even at long distances, under bad weather, and with signal blockage.

  v.   Finally, the proposed system will ensure six significant parameters of the IoV system including data rate, communication range, mobility support, minimum communication delay, and scalability.

## 5.2 Proposed Architecture and Modelling

The proposed intelligent IoV system consists of three main forms of communications: V2V, V2I, and V2C. Fig. 5.2 represents the overall overview of the proposed IoV system based on OCC. For V2V and V2I communication, NIR LEDs, camera are used as transmitter and receiver, respectively. In V2C communication, cellular technology can be used to uphold the connection between the Internet (e.g., cloud or database) and vehicles or infrastructures. The vehicles in V2V communication are defined as forward vehicle and following vehicle. Firstly, the forward vehicle transmits the information and then the following vehicle receives information from the forward vehicle. The vehicles can also receive traffic information (e.g., certain emergency information, including traffic condition, safety information, and accident information) from traffic lights. The forward vehicle uses its back light LEDs (using S2-PSK modulation scheme) to transmit this traffic information towards the following vehicle. The following vehicle uses CNNs to decode the information from the forward vehicles.

In V2C, both forward and following vehicles can share information with cloud server using cellular technology. After receiving information from the vehicles, the cloud server will process the information using a centralized controller (e.g., SDN-based Open Flow [81]) and broadcast this information back to the IoV (i.e., vehicles or traffic lights) network through cellular communication. This information can be useful for the vehicles which are far away from the incident. The details will be explained at the end of this section.

In the remaining parts of this section, the proposed IoV model is described in more details. The proposed system employs a multi-criterion application in the following four phases: (1) detection of NIR LEDs; (2) target achievement; (3) LED-array pattern detection and recognition using CNN; and (4) broadcast information.

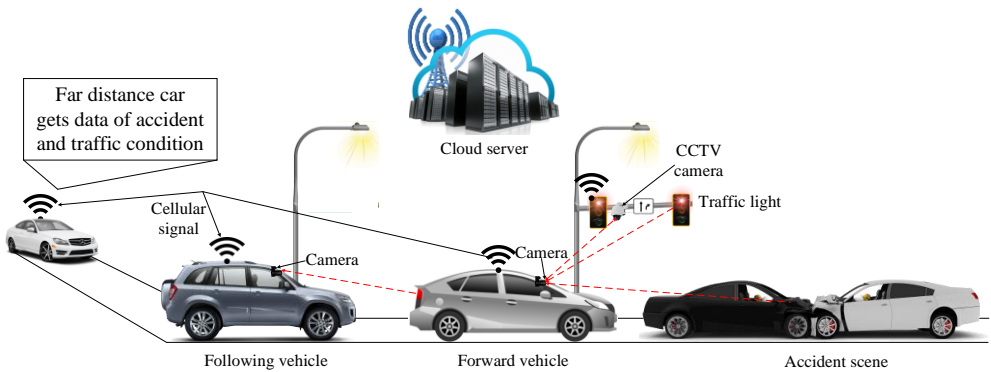

Fig. 5.2 Proposed NIR-based intelligent IoV system architecture with an accident scene

### 5.2.1 Detection of NIR LEDs

The aim of this step is to recognize vehicle NIR LED lights or traffic lights using a camera receiver that captures the whole scene within its FOV. For this case, we considered an example where a vehicle is moving down a road and identifying other vehicles or infrastructures with NIR-optical signals. In NIR-based OCC systems, at the transmitter side, LEDs emit incoherent NIR lights whose intensities are detected by the camera at the receiver end. The emitted signals from transmitters can be detected using IM/DD technique that passes through any point between the transmitter and receiver. For data transmission in OCC, various modulations (such as FSK, PSK, and OOK) have been proposed by the IEEE 802.15.7m standard [35]. Here, we considered S2-PSK modulation scheme [19]. An IS mounted on a vehicle can detect S2-PSK modulated NIR signals.

An overview of NIR LEDs (RoIs) detection process from the image is shown in Fig. 5.3. As indicated in figure, the following vehicle captures an image of a night scenario. The captured images contain not only the NIR LED light sources, but also reflected lights from various surfaces that are not related to the IoV infrastructure. To detect absolute optical signals, high-intensity values in the captured image will be extracted using an image-acquisition process. Generally, the height of the light sources in the road infrastructure is several meters away from the traffic lights depending on the surface of the road. Thus, it is expedient to differentiate between traffic lights and other lights that are mainly used for road decoration or lighting.

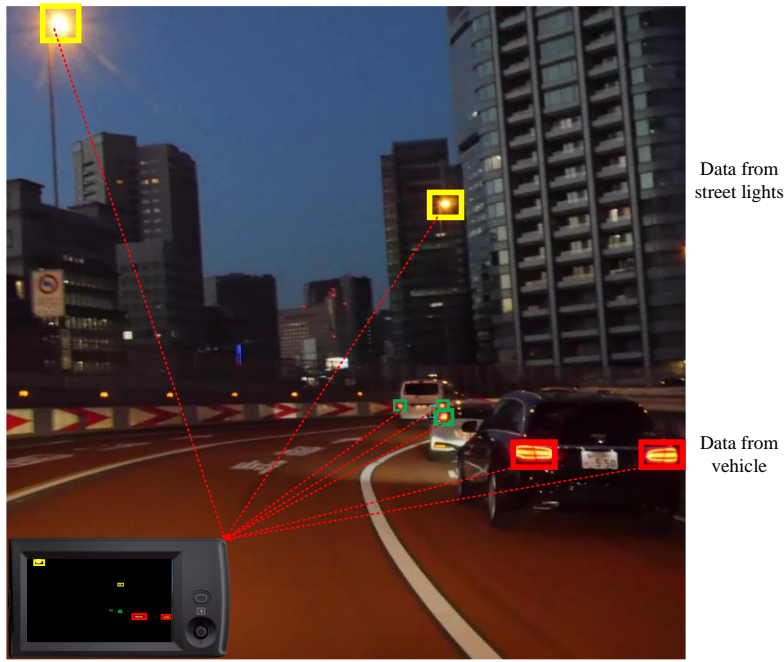

Fig. 5.3 Multiple RoIs detection to identify data sources and remove the interference sources

After the intensity-filtering process, differential images are determined from two consecutively captured images and stored for further processing. The differential images help to recognize the changes between adjacent images. Therefore, the actual NIR LEDs can be distinguished from noise sources in the differential images. Here, the NIR light sources are blinking with S2-PSK modulation which represents two phase "0°" and "180°". Then, the resulting image is binarized to extract the LED signaling features from the captured images.

In summary, the identification algorithm processes the input images to extract the corresponding NIR LED region from the emitters. The other light sources (sunlight, advertising boards) are discarded through shape analysis. The whole identification algorithm can be summarized as follows:

i)  The camera acquires NIR-filtered images in which the emitters appear as bright spots.

ii)  To keep the brightest pixels (NIR LEDs) in the scene, the received image is binarized using a threshold value.

iii) The NIR regions are made more consistent, bright, and less fragmented, using morphological dilatation.

iv)  The regions to be accepted or rejected will mainly depend on the size and shape of the detected region. For example, large regions (corresponding to the sky), regions which don't look like a spot, (such as pieces of vegetation, LED signage, advertising boards), and regions with small spots (such as reflection from other light sources) are rejected. The algorithm used to determine the shape of a region calculates the rate of pixels of the detected region inside its circumcircle.

v)  The accepted spots are termed as RoIs or targets.

### 5.2.2 Distance Measurement

After identifying the LED light sources (i.e., RoIs) on the image plane, the risk factor of action taken by surrounding vehicles is determined (which is also called finding the temporal and spatial condition of a vehicle). For this case, the distance from the captured image is measured. Distance information can help to make decisions in which vehicle information is important and where it is not necessary to pay attention. The distance measurement procedure has been described in 4.2.2.

### 5.2.3 LED-array Pattern Detection and Recognition Using CNN

After obtaining the distance information from target achievement stage, the information from the targeted vehicles are decoded. Here, CNN is used to recognize LED patterns. For instance, if the targeted vehicles are at long distance or LED signals are blocked due to bad weather conditions, other vehicles and infrastructures, it will be difficult to decode the pattern of the LEDs. Though neural networks (NN) and other pattern-recognition algorithms have been developed over the past 50 years, CNNs has developed significantly in the recent past. CNNs are being used in a variety of areas, such as image and pattern recognition, natural-language processing, speech recognition, and video analysis. The improved network structures of CNNs lead to memory savings and reduced computational-complexity and, at the same time, offer better performance for various applications.

Moreover, CNN is robust against distortions, such as different lighting conditions, change in shape due to the camera lens, presence of partial occlusions, horizontal and vertical shifts, and so forth. In the conventional case, using a fully connected layer to extract the features, the input image of size 32x32 and a hidden layer having 1,000 features will require on the order of 106 coefficients, which requires a huge memory. In the CNN layer, the same coefficients are used across different locations in space, so the memory requirement is drastically reduced. Also in the standard NN, the number of parameters is much higher which increases the training time proportionately. On the other hand, the training time is reduced due to the reduced number of parameters. Assuming perfect training, we can design a standard NN similar to CNN according to the performance. In practical case, a NN requires more parameters than a CNN that creates much noise at the time of training. As a result, the performance of a standard NN equivalent to a CNN will always be worse.

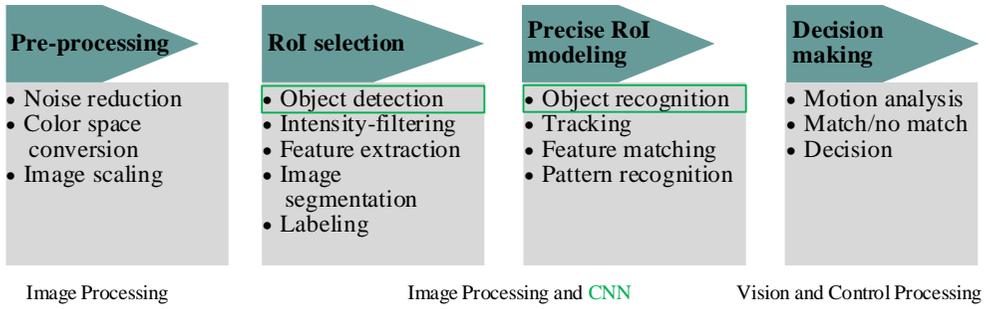

Fig. 5.4 Typical CNN algorithm for detection and recognition

Fig. 5.4 shows a typical CNN algorithm for detection and recognition, which consists of four stages: (i) pre-processing of the image; (ii) detecting the RoI; (iii) object recognition; and (iv) decision making. The first step contains the outside data, which can be used for training, especially the camera parameter. The decision-making step works on the recognized objects; sometimes it may make complex decisions, but it operates on much less data, so these decisions are not usually computationally hard or memory intensive. However, CNNs are now having a wide impact in object-detection and recognition stages, which are the most difficult challenges. Fig. 5.5 shows a visualization of CNN algorithm for a vehicle's LED-pattern-detection and recognition algorithm.

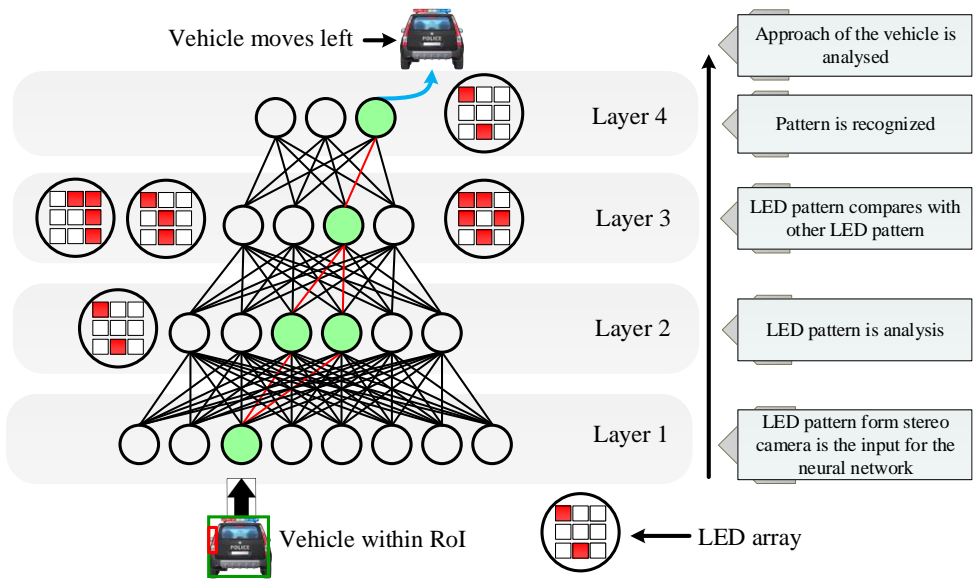

Fig. 5.5 CNN-based LED pattern classification and recognition

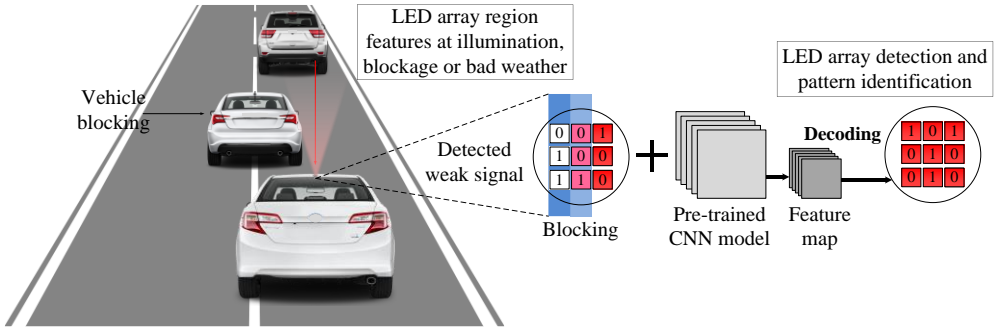

Fig. 5.6 Decoding of LED-array data using CNN at unclear state

Fig. 5.6 shows the operation of LED-state detection using CNN whose signal has been blocked by another vehicle. In our system, CNN is adopted to classify the LED region instead of classifying the whole image, reducing computational cost and total computational time. To better understand the proposed CNN system has been categorized it into four steps, namely design of a CNN, configuration of training options, training a faster CNN object detector, and evaluating the trained detector.

### 5.2.3.1 Designing a CNN

Each layer in CNN performs a definite function. For example, *imageInputLayer* is for image-input layer, *convolutional2dLayer* is for 2D convolution layer in CNN, *reluLayer* is for rectified linear unit (*ReLU*) layer, *fullyConnectedLayer* is for fully connected layer, *maxPooling2dLayer* is for max pooling layer, and *classificationLayer* is for output layer classification and recognition of an NN. The composition of layer-by-layer CNN has done by NN Toolbox™. The first step is to design the input layer, which defines the type and size of the *imageInputLayer* function. The input size varies for different purposes. For classification tasks, the input size is typically the same size as the training images. However, for detection or recognition tasks, the CNN needs to analyze smaller parts (i.e., the LED region) of the image, so the input size must be at least the size of the smallest object in the data set. In this case, the CNN is used to process a [32*32]-RGB image.

Then, the middle layers of the network are made up of convolutional repetitive blocks, *ReLU*, and pooling layers, which are the core part of the CNN. During network training, sets of filter weights in convolution layers are updated; non-linear functions have added to the network *ReLU* layer to map pixel of the image and the pooling layers downsample data as they flow through the network. A deeper network could be generated by repeating these basic layers, but to avoid down-sampling of the data too early, pooling layers should be used cautiously. Because of early down-sampling, the important information for learning can be discarded. All the layers of a CNN are fully interconnected. At this point, the network must produce outputs that can be used to measure whether the input image belongs to one of the object classes or the background. Finally, the three layers were combined. Then, the first convolutional layer weights are initialized with a standard deviation of 0.0001 which improves the convergence of training.

### 5.2.3.2 Configure Training Options

The training of the CNN can be classified into four categories. In the first two categories, the region-proposal and -detection networks are trained. The final two categories combine the networks from the first two steps into a single network [82]. As each training category may have a different convergence rate, each category should be set with independent training options. We can specify the network-training options using the *trainingOptions* function or the Neural Network Toolbox™. In this case, we have set the learning rate for the first two steps higher than for the last two steps, such that the weights can be modified more slowly in the last two steps for fine-tuning purposes. The greatest advantage is that a training can be resumed from a previously saved point even though training is interrupted due to a power outage or system failure.

### 5.2.3.3  Training a CNN LED-pattern-recognition Detector

After specifying the CNN and training options, we need to train the LED-pattern detector. The input of this detector is the pre-trained network and the training options. The training function can form a new network by modifying the original trained network automatically. The image patterns (i.e., LED patterns) are extracted from the training data during this process. The patterns required for training are defined by "*PositiveOverlapRange*" and "*NegativeOverlapRange*". Positive training samples overlap by 0.6 to 1.0, whereas negative training samples overlap by 0 to 0.3. The best values for the positive-negative pairs should be chosen based on the testing value of the training detector.

To accelerate CNN training and reduce the training time, the use of a parallel pool is highly recommended for MATLAB users. But the parallel pool should be enabled prior to training. For computational competence, a GPU of 3.0 or higher is strongly recommended. To save execution time, a pre-trained network can be loaded from a disk. If one desires to train the network oneself, one must set the *doTraining* variable manually.

### 5.2.3.4  Evaluating the Detector Using a Test Set

To verify the training, the detector investigation for a test image is required. The primary step for detector performance evaluation is to collect the detection results by running the detector on a test image set. To ensure a short evaluation time, the results are loaded from a previously saved disk. For this, the *doTraining* function was set from the previous section to execute the evaluation locally. To evaluate the detector effectively, it is recommended to test larger image sets. The common performance metrics can be measured using object-detector-evaluation functions, which are supported by the MATLAB Computer Vision System Toolbox™; for example, log-average miss rates can be found using the *evaluateDetectionMissRate* function and average precision can be found using *evaluateDetectionPrecision* function.

### 5.2.4 Broadcast Information Using a Central Server

In the proposed scheme, the concept of cloud server or database for non-LOS (NLOS) communication have been introduced. After receiving the information using CNN, emergency information (e.g., accidents or traffic condition) will be broadcasted to the cloud server in order to support communication with distant vehicles. Suppose, an emergency condition (e.g., accident) has occurred far (5 km) up the road from the host vehicle. If the remote vehicles can get this information instantly, it will be easy to change their route according to the traffic conditions. However, to broadcast this information to the remote vehicles instantly using OCC-based V2V communication, it will be time-consuming to reach long distances (e.g., 5 km). Thus, in this case, OCC-based communication will not be effective. As a result, cloud-based vehicular communication for long distances has been introduced.

In V2C communication, the vehicles, at the incident can receive the information from the forward vehicles using cameras. Then, the processing system mounted on the vehicles will transmit the emergency information to the cloud server using cellular. After receiving information from the vehicles, the cloud server will process the information using a centralized controller (e.g., SDN-based Open Flow [81]) and broadcast the information back to the IoV networks based on the priority of the incident to all connected links (i.e., vehicles or traffic lights) through cellular technology. After receiving the information from the server, the vehicles or traffic lights will transmit that information through NIR LED lights to the following vehicles, allowing them to change direction or take other actions to reach their destinations based on the situation. However, this is out of the scope of this paper.

#### 5.2.4.1 SDN for IoV

Networking technologies have experienced explosive evolution during the last decade. Moreover, the number of mobile devices and the data traffic are increasing exponentially because network applications are extending from the traditional hardware-based to real-time communication in social networks, e-commerce, and entertainment. However, hardware-based network systems mainly depend on insecure and inflexible network architectures which generally take a typical 10-year for a new generation of wireless networks to be standardized and deployed. To facilitate these challenges of current network architectures, the most important task is to shift the design of current architectures for the next-generation wireless networks. Moreover, the complementary concept of SDN, NFV has been presented to separate the control functionalities from the hardware by simply decoupling the forwarding plane from the control plane. These functionalities will ensure the required flexibilities and adaptability of the ever-changing network architectures with the introduction of the concept of SDN.

SDN has been introduced for data networks and next-generation Internet [83]. SDN has the following characteristics: (i) it decouples network control from the underlying data plane (e.g., routers, vehicle node, switches,); (ii) it allows the control plane to be programmed directly through an open interface, for instance, OpenFlow [84]; and (iii) it uses a network controller to define the behavior and operation of the networking infrastructure (i.e., SDN controller). SDN can be an ideal prospective for the high-bandwidth, dynamic nature of network management. SDN provides the flexibility to change the network configuration at the software level, thus reducing the necessity of modification at the hardware level. SDN makes it easier to introduce and deploy new applications and services than the traditional hardware-operated networking architectures. It also ensures the quality-of-service (QoS) at any level of user requirements. Consequently, it will be an enticing architecture from the viewpoint of reconfiguring and redirecting complex networks for real-time management.

An important observation of SDN is NFV [85]. SDN and NFV are mutually beneficial, but they are not fully dependent on each other. In fact, network functions can be employed and virtualized without using an SDN and vice versa. As it is complementary to SDN, NFV can effectively decouple network functionalities and implement them in software. Thus, it can decouple network functions, for instance; routing decisions, from the underlying hardware devices such as routers and switches, and centralize them at remote network servers or in the cloud through an open interface such as OpenFlow. Hence, the overall network architecture can be highly flexible for fast and adaptive reconfiguration.

### 5.2.4.2 Proposed Architecture

Open networking foundation (ONF) proposed a simple high-level architecture for SDN. This model can be separated into three layers, namely, an infrastructure layer, a control layer, and an application layer, assembled over each other. The details these three layers are described below. Fig. 5.7 illustrates the SDN architecture for the IoV system.

### a) Infrastructure Layer

The infrastructure layer mainly consists of forwarding elements (e.g., physical and virtual switches, routers, wireless access points, and etc.) that comprise the data plane. These devices are mainly responsible for (i) collecting network status, storing

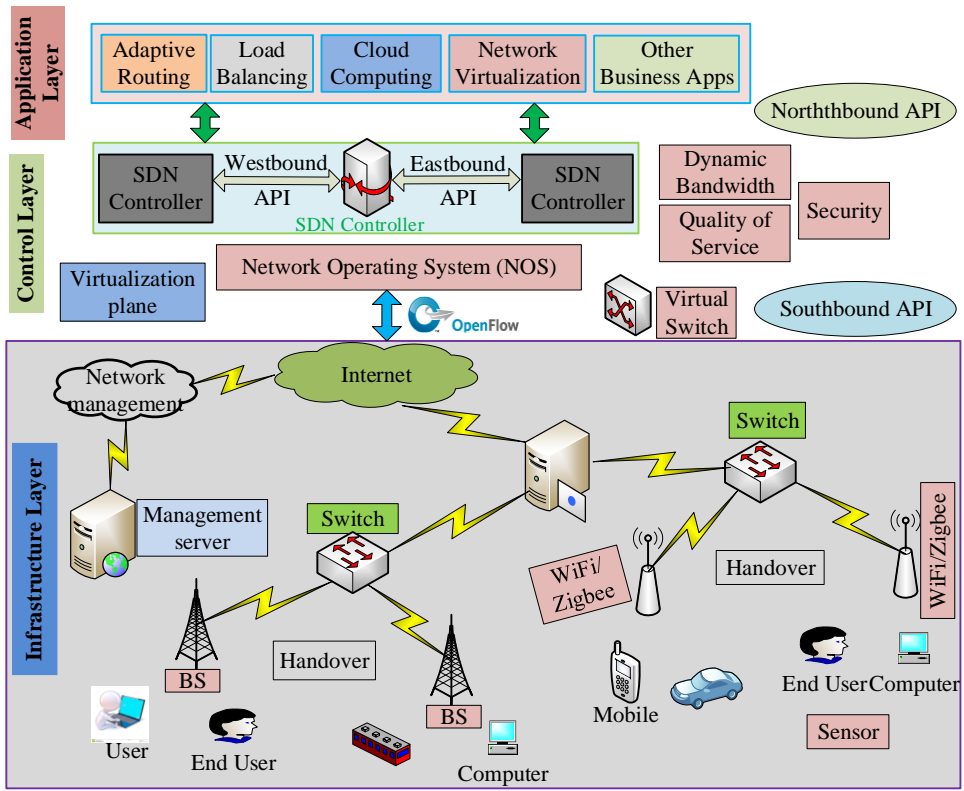

Fig. 5.7 SDN architecture for IoV

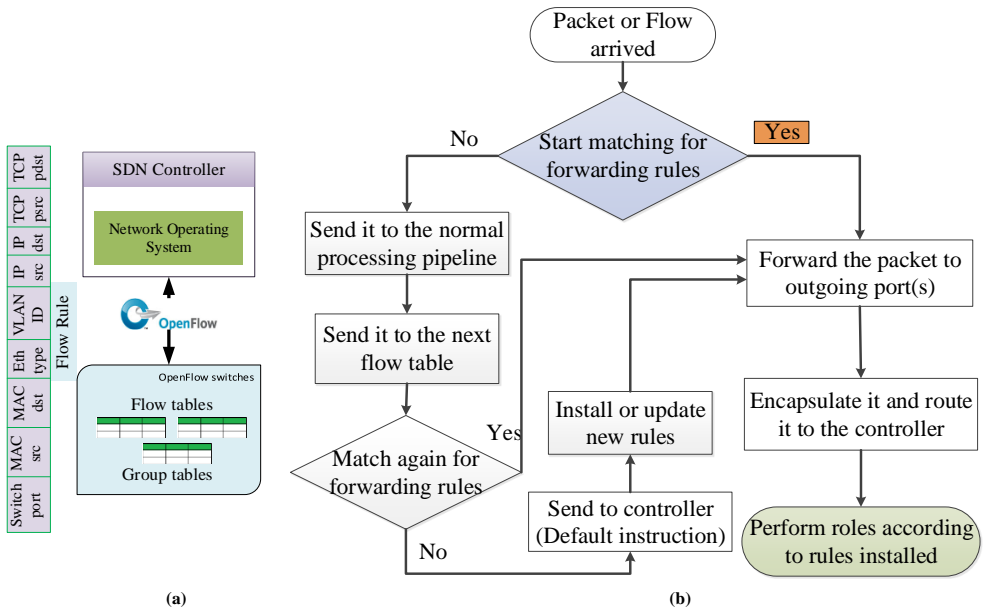

Fig. 5.8 (a) OpenFlow model and (b) detailed process of OpenFlow protocol them temporally in local network devices and sending the stored data to the network controllers and (ii) for managing packets based on the rules provided by the network controllers or administrators. They allow the SDN architecture to perform packet switching and forwarding via an open interface known as the southbound interface. In most SDNs, OpenFlow is used as the open southbound interface. OpenFlow uses flow-oriented protocol. It has port and switches and port for flow control.

**OpenFlow:** ONF is a fundamental component for developing SDN solutions and possibly treated as an encouraging consideration for any networking abstraction. Main function of ONF is to maintain the OpenFlow protocol. OpenFlow helps SDN architecture to adjust the dynamic of user applications, high-bandwidth, adjust the network to different business needs, and decrease maintenance and management complexity. Fig. 5.8(a) shows the model of the OpenFlow protocol whereas the algorithm is shown in the Fig. 5.8(b). As shown in figure, when a new flow or packet reaches, some lookup manner originates in the primary lookup table and concludes either with a match in the flow tables or with an error depending on the rules specified by the controller. A delinquency info has forwarded to the controller, if the packet is sent without acknowledgement of very next step "'send to controller" in the case of any unmatched entry. Event based message such as triggering port or link change are sent to the controller by forwarding devices. Once the rules are matched with the flow rules, the rule's counter is incremented and actions based on the set rules start getting executed. This could lead to forwarding of a packet, after modifying some of its header fields to a specific port or (i) dropping of the packet, (ii) reporting of the packet back to the controller. However, there have other API proposals [81] rather than OpenFlow as southbound interface.

### b) Network Controller

The network controller, network operating system (NOS) or SDN controller, is the heart of SDN architecture. It lies between network devices and applications. It is based on operating systems in computing. In [86], the controller is described as software abstraction that controls all functionalities of any networking system. It maintains control over the network through three interfaces, namely, southbound interface (e.g., OpenFlow), northbound interface (e.g., API), and east/westbound interfaces. The southbound interface abstracts the functionalities of programmable switches and connects them to the controller. The northbound interface allows high-level policies or network applications to be deployed easily and transmits them to the NOS, while the east/westbound interfaces maintain communications between groups of SDN controllers. Thus far, many SDN controllers have been proposed by researchers to facilitate controller functionalities [81].

The SDN controller **functionalities** can be divided into four types: (i) a high-level language to describe network actions; (ii) a rule update to put in regulation generated from high-level languages; (iii) a network status collection process to gather network infrastructure information; and (iv) a network status synchronization process to build a global network view using the network statuses collected by each individual controller [81].

### c) Application Layer

At the top of SDN architecture, the application layer is located (Fig. 5.7). SDN application interacts using northbound interface with the controller to achieve an unambiguous network function. They request network services or user requirements and then manipulate these services. Although there is a well-defined standardized southbound interface such as OpenFlow, there is no standard northbound interface for interactions between controllers and SDN applications. Therefore, we can say that the northbound interface is a set of software-defined APIs, not a protocol. SDN applications can provide a global network view with instantaneous status through northbound APIs.

### 5.2.4.3   Challenges and Future Direction

Though we presented the advantages of SDN for 5G, SDN has confronted with some challenges. First of all, security is a more challenging task that needs to be available everywhere within the SDN architecture because of: i) architecture and its controller, applications, devices, channels (TLS with plain text) and flow table, ii) connected resources, iii) services (to protect availability), and iv) information. Furthermore, a reliable and balance controller is still in the out of scope because of lacking of robust and reliable framework policy. The framework policy should be very simple to maintain and implement, secure, and cost effective.

As SDN is a fully automated system with a centralized controller that has reducing human control, error free and fast configuration [87]. Despite these challenges, some remaining implementation affairs need to acknowledge such as flow tables and their large number of flow entries, flow level programming and controller programming, flow instructions and actions. Though NFV and SDN are independent and complementary to each other, they can provide an open environment to fasten the innovation, and can be easily integrated with new services and infrastructure like controlled and automated network resources. Packet forwarding or processing is performed by NFV, while the controller can control or update flow tables according to the needs of users or applications at any level. Here, NFV is responsible for creating or processing flow rules, and SDN is responsible for the management of the said rules. Integration of SDN and NFV will be a promising technology for future IoV.

## 5.3 Simulation Results

The first part of the proposed algorithm is to detect the RoI (i.e., the NIR LED region in the image) where a real road video to detect the RoIs on the image plane. Fig. 5.9(a) shows an image frame of a time lapse video and Fig. 5.9(b) represents the detected RoIs on the image plane after applying the algorithm. In this case, the NIR LED regions were extracted using the differential images and then the resulting image is binarized using RGB thresholding. In the figure, the detected RoI with red rectangular markings represents the nearest vehicle, the green rectangular markings represent the far-distant vehicles, and the yellow rectangular markings represent the signals from the traffic lights. Fig. 5.9(c) and (d) show 3D representations of the original and optimized constellations of the threshold image at the receiver. It can

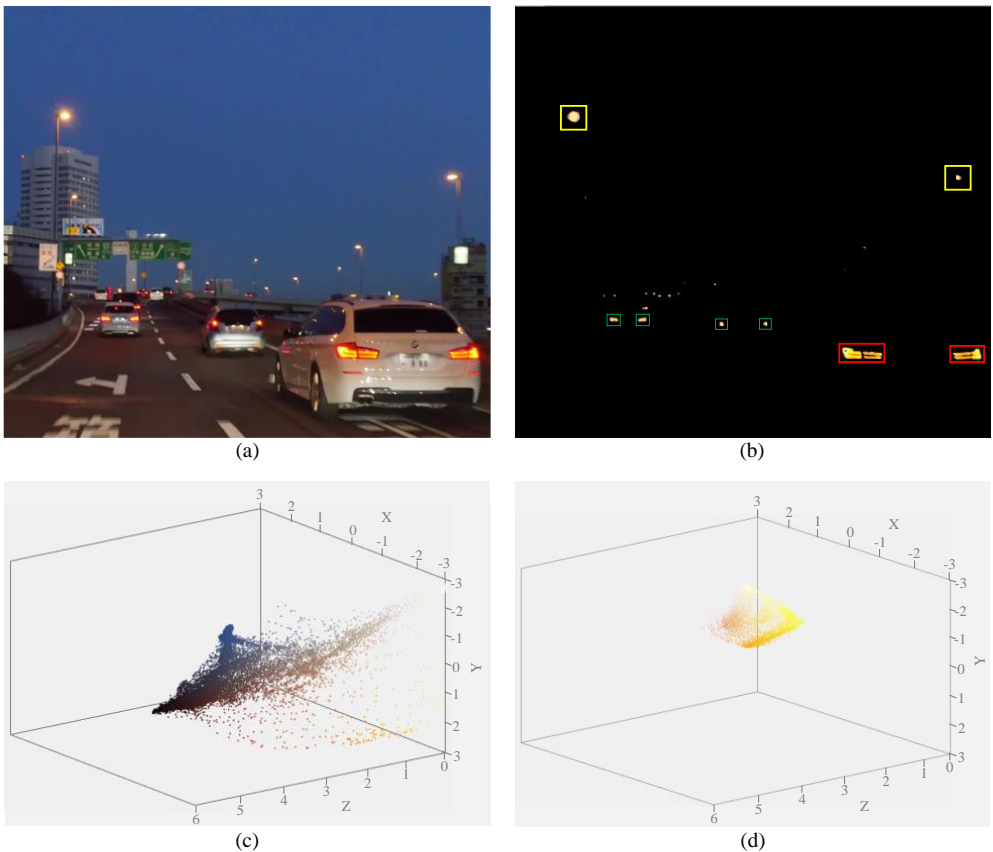

(a)  (b)

(c)  (d)

Fig. 5.9 (a) Original image, (b) successful detection of NIR LED (i.e., RoI) and 3D thresholding of image at the receiver end (c) before optimization, (d) after optimization be seen that the optimized constellation points in 3D color space are distributed more uniformly than in un-optimized or original 3D color space. It implies that most of the constellation points are close to the threshold value, which proves the high efficiency of our proposed algorithm. But there have a few abnormal constellation points are away from the threshold-constellation points, which can be ignored.

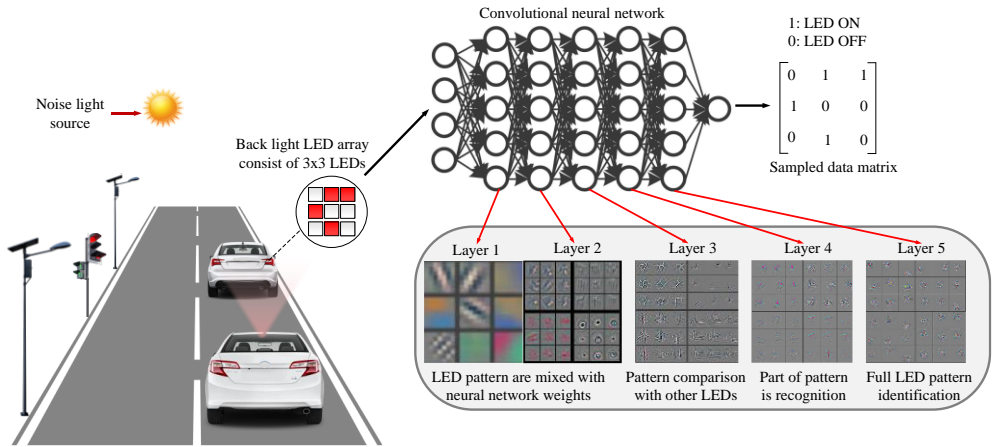

Fig. 5.10 CNN-based LED-light classification scenario under ambient-noise lights condition

Fig. 5.10 presents an example of how a CNN-based system can easily distinguish desired lights information from ambient or others noise sources. The detection and recognition performance were determined by measuring the average precision and average miss rate of our CNN algorithm. In the first step, the previous detection results were utilized by running the detector on the test-image sets to evaluate detector-performance and then, executed the detection and recognition algorithm on an intended scenario. To avoid long evaluation times, the results were loaded from disk and the *doTraining* MATLAB function was set to execute the algorithm locally. Table 5.1 represents the performance parameters used for the execution of the CNN-based detection algorithm. It shows that weak learners in the trained classifier decrease in the final stage.

Table 5.1 Performance parameters for CNN-based object detection

| Object detector training | 3 Stages |
|---|---|
| Model size | 34x31 |
| Stage 1: (42 positive examples and 210 negative examples) | The trained classifier has 87 weak learners |
| Stage 2: (Found new 210 negative examples) | The trained classifier has 51 weak learners |
| Stage 3: (42 positive examples and 210 negative examples) | The trained classifier has 19 weak learners |
| Elapsed time (seconds) | 13.3242 for precision rate and 13.202 for miss rate |

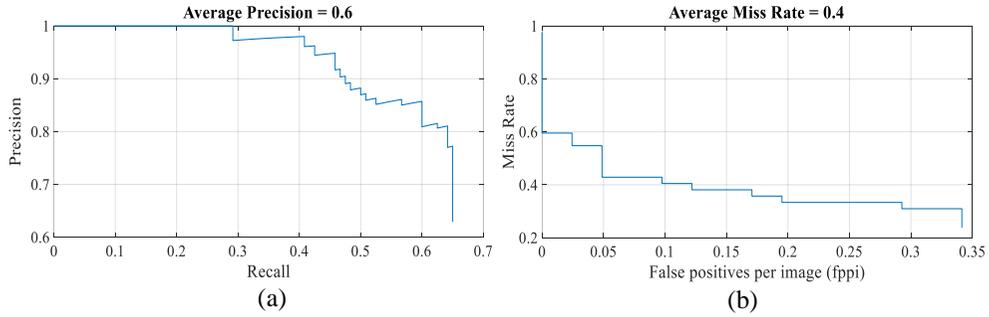

Fig. 5.11 Performance of the object detector: (a) average precision and (b) average miss rate

300 vehicle samples were trained for the CNN system of which 60% of the data were used for training to find the precise detection and miss rates. Fig. 5.11 compares the detection and miss rates using our proposed CNN-based detection and recognition algorithm. The precision/recall (PR) curve highlights how precise the detector is at varying levels of recall. The average precision provides a single number that incorporates the ability of the detector to make correct classifications (precision) and the ability of the detector to find all relevant objects (recall). Ideally, the precision would be 1 at all recall levels. In this example, the average precision is 0.6 and the average miss rate is 0.4. The use of additional layers in the network can help to improve the average precision but might require additional training data and longer training times. In this example, the proposed approach used a single image, which showed promising results. To fully evaluate the detector, testing it on a larger set of images is recommended.

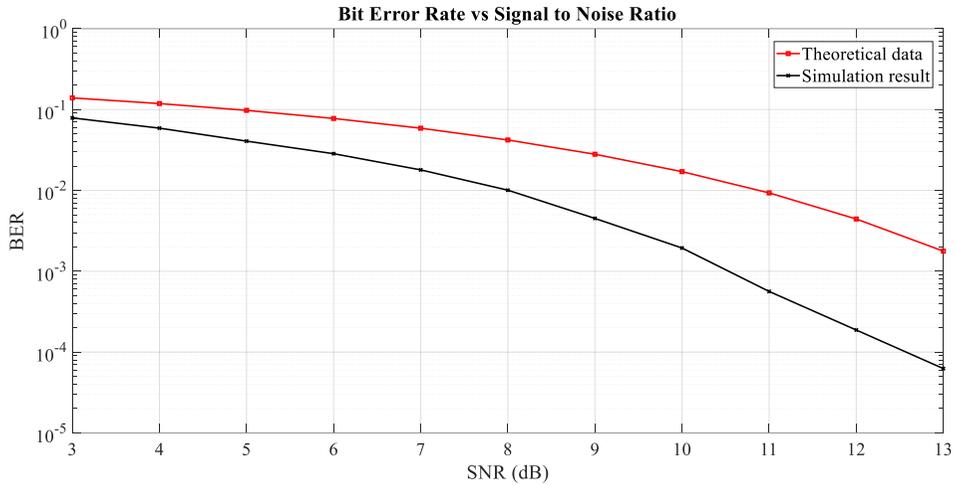

Fig. 5.12 SNR vs BER performance for CNN-based IoV system

Meanwhile, BER is the number of bit errors in the number of received bits over the communication channel due to noise, interference, distortion, or bit-synchronization errors. Therefore, minimizing BER is an important issue for maintaining good communication between multiple vehicles. The desired results of BER with respect to SNR are obtained for the optical channel (considering AWGN in the channel). Simulation result for optical channel has shown good performance when it is compared to the theoretical value (see Fig. 5.12). In other words, the system can show poor performance when the number of vehicles is increased.

## 5.4 Conclusion

IoV has emerged as a new technology due to the revolution of automated vehicles in the recent years. Many technologies have been proposed to support IoV. Due to the limitations of the existing technologies NIR based OCC system has been introduced in this part of the thesis. The NIR based IoV system comprised of NIR LEDs which act as transmitters and cameras which act as receivers mainly focusing on automotive vehicular applications. Here, the LED lights were detected by discarding other light sources (i.e., sky, digital displays, advertising boards) using differential images and thresholding value. A real-time video was used to obtain the simulation results. The results indicate the efficiency of the algorithm to differentiate NIR LEDs from other light sources. We then applied the stereo camera-based distance-measurement algorithm to find the distance between the forward and following vehicle. It helps the following vehicle to decode information from the forward vehicle. To verify the stability of the system, the mean re-projection errors were calculated during the camera-calibration process using several image pairs. After getting the distance value from the stereo-vision process, CNN was applied to detect and recognize the LED array pattern form the desired targets or RoIs. The results using CNN show that the algorithm can recognize the LED array pattern preciously, even under signal-blockage and bad-weather conditions.

# Chapter 6
# Conclusions and Discussions

## 6.1 Summary

Optical camera communication (OCC) has been considered as a supportive and efficient candidate for effective AV communication. It has proved it's applications in next generation communication providing communication, identification, and illumination simultaneously. To provide contribution in the upcoming research, an intelligent OCC system has been proposed for IoV system to provide smart vehicular communication. Firstly, vision camera and high-speed camera based approach has been proposed for multiple vehicles detection and fast data processing simultaneously. Furthermore, NIR based OCC has been introduced to provide communication at bad weather condition. To decode information at bad weather condition, CNN has been proposed. The use of high-speed camera and CNN provide better performance than the conventional computer vision system.

## 6.2 Future Works

CNN is sometime time consuming and complicated to implement. So, some algorithm will be used to reduce the time and complexity considering the significant drawbacks of the proposed scheme. Moreover, the implementation work for the outdoor environment will be commenced in future. However, OCC can be connected to another network through IoT functionality. This issue will be included in the future work. In the future, the mobility support of the transmitter and receiver will be considered.

# References

[1] "SBD: Connected car global forecast," 2015. [Online]. Available: https://www.sbdautomotive.com/files/sbd/pdfs/536%20connected%20car%20forecast%20ib%20 15.pdf.

[2] F. Yang, S. Wang, J. Li, Z. Liu, and Q. Sun, "An overview of Internet of Vehicles," *China Communications,* vol. 11, no. 10, pp. 1-15, Oct. 2014.

[3] A. Dua, N. Kumar, and S. Bawa, "A systematic review on routing protocols for vehicular Ad Hoc," *Vehicular Communications,* vol. 1, no. 1, pp. 33-52, 2014.

[4] M. Saini, A. Alelaiwi, and E. Saddik, "How close are we to realizing a pragmatic VANET solution? A meta-survey," *ACM Computing Survey,* vol. 48, no. 2, pp. 1-40, 2015.

[5] J. Barbaresso, G. Cordahi, D. E. Garcia, C. Hill, A. Jendzejec, and K. Wright, "USDOT's Intelligent transportation systems (ITS) ITS strategic plan 2015–2019," US Department of Transportation, Intelligent Transportation Systems, Joint Program Office, Washington, DC, USA, 2015.

[6] B. W. Parkinson and J. J. Spilker, Global Positioning System: Theory and Application, American Institute of Astronautics and Aeronautics, 1996.

[7] A. Islam, M. A. Hossain, and Y. M. Jang, "Interference mitigation technique for time-of-flight (ToF) camera," in *2016 Eighth International Conference on Ubiquitous and Future Networks (ICUFN)*, Vienna, 2016.

[8] A. Joshi and M. R. James, "Generation of accurate lane-level maps from coarse prior maps and LIDAR," *IEEE Intelligent Transportation Systems Magazine,* vol. 7, no. 1, pp. 19-29, 2015.

[9] M. A. Hossain, A. Islam, N. T. Le, H. W. Lee, Y. T. Lee, and Y. M. Jang, "Performance analysis of smart digital signage system based on software-defined IoT and invisible image sensor communication," *International Journal of Distributed Sensor Networks,* vol. 12, no. 7, pp. 1-14, Jul. 2016.

[10] T. Yamazato et al., "Image-sensor-based visible light communication for automotive applications," *IEEE Communications Magazine,* vol. 57, no. 7, pp. 88-97, Jul. 2014.

[11] I. Takai, S. Ito, K. Yasutomi, K. Kagawa, M. Andoh, and S. Kawahito, "LED and CMOS image sensor based optical wireless communication system for automotive applications," *IEEE Photonics Journal,* vol. 5, no. 5, pp. 6801418-6801418, Oct. 2013.

[12] T. Komine, and M. Nakagawa, "Fundamental analysis for visible-light communication system using LED lightings," *IEEE Transactions on Consumer Electronics,* vol. 50, no. 1, pp. 100-107, 2004.

[13] D. C. O'Brien, L. Zeng, H. Le-Minh, G. Faulkner, J. W. Walewski and S. Randel, "Visible light communications: challenges and possibilities," in *International Symposium on Personal, Indoor and Mobile Radio Communications*, Cannes, 2008.

[14] Y. U. Lee and M. Kavehrad, "Long-range indoor hybrid localization system design with visible light communications and wireless network," in *2012 IEEE Photonics Society Summer Topical Meeting Series*, Seattle, WA, 2012.

[15] S. Y. Jung, S. Hann, and C. S. Park, "TDOA-based optical wireless indoor localization using

LED ceiling lamps," *IEEE Transactions on Consumer Electronics,* vol. 57, no. 4, pp. 1592-1597, Nov. 2011.

[16] S. Okada, T. Yendo, T. Yamazato, T. Fujii, M. Tanimoto, and Y. Kimura, "On-vehicle receiver for distant visible light road-to-vehicle communication," in *IEEE Intelligent Vehicles Symposium*, 1033–1038, Jun. 2009.

[17] A. Cailean, B. Cagneau, L. Chassagne, S. Topsu, Y. Alayli, and J. M. Blosseville, "Visible light communications: Application to cooperation between vehicle and road infrastructures," in *Intelligent Vehicles Symposium (IV)*, Jun. 2012.

[18] T. Yamazato et al, "Vehicle motion and pixel illumination modeling for image sensor based visible light communication," *IEEE Journal on Selected Areas in Communications,* vol. 33, no. 9, p. 1793–1805, Sep. 2015.

[19] T. Nguyen, A. Islam, and Y. M. Jang, "Region-of-interest signaling vehicular system using optical camera communications," *IEEE Photonics Journal,* vol. 9, no. 1, pp. 1-20, Feb. 2017.

[20] S. A. Bagloee, M. Tavana, M. Asadi, and T. Oliver, "Autonomous vehicles: challenges, opportunities, and future implications for transportation policies," *Journal of Modern Transportation,* vol. 24, no. 4, p. 284–303, Dec. 2016.

[21] J. Nilsson, A. C. E. Ödblom, and J. Fredriksson, "Worst-case analysis of automotive collision avoidance systems," *IEEE Transactions on Vehicular Technology,* vol. 65, no. 4, pp. 1899-1911, Apr. 2016.

[22] J. Guo, P. Hu, and R. Wang, "Nonlinear coordinated steering and braking control of vision-based autonomous vehicles in emergency obstacle avoidance," *IEEE Transactions on Intelligent Transportation Systems,* vol. 17, no. 11, pp. 3230-3240, Nov. 2016.

[23] B. Wang, S. A. R. Florez, and V. Frémont, "Multiple obstacle detection and tracking using stereo vision: application and analysis," in *International Conference on Control Automation Robotics & Vision (ICARCV)*, 2014.

[24] Y. C. Lim, J. Kim, C. H. Lee, and M. Lee, "Stereo-based tracking-by-multiple hypotheses framework for multiple vehicle detection and tracking," *International Journal of Advanced Robotic Systems,* vol. 10, no. 7, Jul. 2013.

[25] M. Qing, and Kang-Hyun Jo, "A novel particle filter implementation for a multiple-vehicle detection and tracking system using tail light segmentation," *International Journal of Control, Automation and Systems,* vol. 11, no. 3, p. 577–585, Jun. 2013.

[26] N. E. Faouzi, H. Leung, and A. Kurian, "Data fusion in intelligent transportation systems: progress and challenges – a survey," *Information Fusion,* vol. 12, no. 1, p. 4–10, Jan. 2011.

[27] I. Takai, T. Harada, M. Andoh, K. Yasutomi, K. Kagawa, and S. Kawahito, "Optical vehicle-to-vehicle communication system using LED transmitter and camera receiver," *IEEE Photonics Journal,* vol. 6, no. 5, pp. 1-14, Oct. 2014.

[28] Y. Goto et al., "A new automotive VLC system using optical communication image sensor," *IEEE Photonics Journal,* vol. 8, no. 3, pp. 1-17, Jun. 2016.

[29] S. Al-Sultan, M. M. Al-Doori, A. H. Al-Bayatti, and H. Zedan,, "A comprehensive survey on vehicular Ad Hoc network," *Journal of Network and Computer Applications,* vol. 37, no. 1, p. 380–392, Jan. 2014.

[30] S. F. Hasan, X. Ding, N. H. Siddique, and S. Chakraborty, "Measuring disruption in vehicular communications," *IEEE Transactions on Vehicular Technology,* vol. 60, no. 1, pp. 148-159, 2011.

[31] J. Toutouh and E. Alba, "Light commodity devices for building vehicular ad hoc networks: An experimental study," *Ad Hoc Networks,* vol. 37, no. 1, pp. 499-511, Feb. 2016.

[32] B. Aslam, P. Wang, and C. C. Zou, "Extension of Internet access to VANET via satellite receive-only terminals," *International Journal of Ad Hoc and Ubiquitous Computing,* vol. 14, no. 3, pp. 172-190, Dec. 2013.

[33] S. Bitam, A. Mellouk, and S. Zeadally, "VANET-cloud: A generic cloud computing model for vehicular ad hoc networks," *IEEE Wireless Communication,* vol. 22, no. 1, pp. 96-102, 2015.

[34] "Visible Light Communication Consortium," [Online]. Available: www.vlcc.net.

[35] "Official website of IEEE 802.15.7m," IEEE, [Online]. Available. https://mentor.ieee.org/802.15/documents?is_dcn=DCN%2C%20Title%2C%20Author%20or%20Affiliation&is_group=007a.

[36] "IG VAT," IEEE 802.15, [Online]. Available. https://mentor.ieee.org/802.15/documents?is_dcn=DCN%2C%20Title%2C%20Author%20or%20Affiliation&is_group=0vat.

[37] "Standards catalogue," ISO, [Online]. Available. https://www.iso.org/committee/54706/x/catalogue/.

[38] M. S. Ifthekhar, N. T. Le, M. A. Hossain, T. Nguyen, and Y. M. Jang, "Neural network-based indoor positioning using virtual projective invariants," *Wireless Personal Communications,* vol. 86, no. 4, pp. 1813-1828, 2016.

[39] M. S. Z. Sarker et al., "A CMOS imager and 2-D light pulse receiver array for spatial optical communication," in *IEEE Asian Solid-State Circuits Conference*, Nov. 2009.

[40] A. Islam, M. A. Hossain, T. Nguyen, and Y. M. Jang, "High temporal-spatial resolution optical wireless communication technique using image sensor," in *2016 International Conference on Information and Communication Technology Convergence (ICTC)*, Jeju Island, South Korea, 2016.

[41] D. Ionescu et al., "A 3D NIR camera for gesture control of video game consoles," in *2014 IEEE International Conference on Computational Intelligence and Virtual Environments for Measurement Systems and Applications (CIVEMSA)*, Ottawa, ON, 2014.

[42] S.D. Perli, N. Ahmed, and D. Katabi, "PixNet: Interference-free wireless links using LCD-camera pairs," in *MobiCom'10, ACM*, New York, USA, 2010.

[43] T. Hao, R. Zhou, and G. Xing, "COBRA: Color Barcode Streaming for Smartphone Systems," in *MobiSys'12*, New York, 2012.

[44] S. H. Chen and C. W. Chow, "Color-shift keying and code-division multiple-access transmission for RGB-LED visible light communications using mobile phone camera," *IEEE Photonics Journal,* vol. 6, no. 6, pp. 1-6, Dec. 2014.

[45] Nobuo Iizuka, "CASIO Response to 15.7r1 CFA," document no. 15-15-0173r1," Mar. 2015. [Online]. Available: https://mentor.ieee.org/802.15/dcn/15/15-15-0173-01-007a-casio-response-to-15-7r1-cfa.pdf.

[46] S. H. Chen, C. W. Chow, "Hierarchical scheme for detecting the rotating MIMO transmission of the in-door RGB-LED visible light wireless communications using mobile phone camera," *Opt. Communication,* vol. 335, pp. 189-193, Jan. 2015.

[47] H. G. Q. P. W. Hu, "LightSync: Unsynchronized visual communication over screen-camera links," in *19th Annual International Conference on Mobile Computing & Networking*, New York,

USA, Oct. 2013.

[48] Y. Long, S. Wang, W. Wu, X. Yang, G. Jeon, and K. Liu, "Structured-light-assisted wireless digital optical communications," *Optics Communication,* vol. 355, p. 406–410, Nov. 2015.

[49] H. S. Liu and Pang G, "Positioning beacon system using digital camera and LEDs," *IEEE Transactions on Vehicular Technology,* vol. 52, no. 2, pp. 409-419, Mar. 2003.

[50] K. Ebihara, K. Kamakura, and T. Yamazato, "Spatially-modulated space-time coding in visible light communications using 2×2 LED array," in *IEEE Asia Pacific Conference on Circuits and Systems (APCCAS)*, Ishigaki, 2014.

[51] K. Ebihara, K. Kamakura and T. Yamazato, "Layered transmission of space-time coded signals for image-sensor-based visible light communications," *Journal of Lightwave Technology,* vol. 33, no. 20, pp. 4193-4206, Oct. 2015.

[52] T. Nagura, T. Yamazato, M. Katayama, T. Yendo, T. Fujii, and H. Okada, "Improved Decoding Methods of Visible Light Communication System for ITS Using LED Array and High-Speed Camera," in *IEEE Vehicular Technology Conference (VTC)*, Taipei, 2010.

[53] H. Okada, T. Ishizaki, T. Yamazato, T. Yendo, and T. Fujii, "Erasure coding for road-to-vehicle visible light communication systems," in *IEEE Consumer Communications and Networking Conference (CCNC)*, Las Vegas, NV, 2011.

[54] R.D. Roberts, "Undersampled frequency shift ON-OFF keying (UFSOOK) for camera communications (CamCom)," in *Wirel. and Optical Commun. Conf.*, 2013.

[55] P. Luo et al., "Experimental Demonstration of RGB LED-Based Optical Camera Communications," *IEEE Photonics Journal,* vol. 7, no. 5, pp. 1-12, Oct. 2015.

[56] C. Danakis, M. Afgani, G. Povey, I. Underwood, H. Haas, "Using a CMOS camera sensor for visible light communication," in *IEEE Globecom Workshops*, 2012.

[57] Y. Liu et al, "Light encryption scheme using light-emitting diode and camera image sensor," *IEEE Photonics Journal,* vol. 8, no. 1, pp. 1-7, Feb. 2016.

[58] T. Nguyen, M. A. Hossain, and Y. M. Jang, "Design and implementation of a novel compatible encoding scheme in the time domain for image sensor communication," *Sensors,* vol. 16, no. 5, May 2016.

[59] H. Aoyama and M. Oshima, "Visible light communication using a conventional image sensor," in *IEEE Consumer Communications and Networking Conference (CCNC)*, Las Vegas, 2015.

[60] N. Rajagopal, P. Lazik, and A. Rowe, "Visual light landmarks for mobile devices," in *International Symposium on Information Processing in Sensor Networks (IPSN)*, Berlin, 2014.

[61] H. Y. Lee, H.M. Lin, Y. L. Wei, H.I Wu, H. M. Tsai, and K. C. J. Lin, "RollingLight: Enabling Line-of-Sight Light-to-Camera Communications," in *International Conference on Mobile Systems, Applications, and Services*, Florence, Italy, May 2015.

[62] C. H. Hong, T. Nguyen, N. T. Le, and Y. M. Jang, "Modulation and Coding Scheme (MCS) for Indoor Image Sensor Communication system," *Wireless Personal Communications,* pp. 1-17, Feb. 2017.

[63] N. Iizuka, "OCC proposal of scope of standardization and applications," IEEE 802.15 SG7a standardization documents, 2014.

[64] T. Nguyen, N.T. Le, Y.M. Jang, "Asynchronous scheme for unidirectional optical camera communications (OCC)," in *Int. Conf. Ubiquitous and Future Net.*, 2014.

[65] H.-Y. LEE, "Unsynchronized visible light Communications using rolling shutter camera: implementation and evaluation," 2014.

[66] S. J. Han, Y. J Han, and H. S. Hahn, "Vehicle detection method using harr-like feature on real time system," *World Academy of Science, Engineering and Technology,* pp. 455-459, 2009.

[67] Z. Sun, R. Miller, G. Bebis, and D. DiMeo, "A real-time precrash vehicle detection system," in *Workshop on Applications of Computer Vision*, 2002.

[68] C. Mei and P. Rives, "Single view point omnidirectional camera calibration from planar," in *IEEE International Conference on Intelligent Robots and Systems (IROS 2007)*, Apr. 2007.

[69] D. Scaramuzza, A. Martinelli, and R. Siegwart, "A toolbox for easily calibrating omnidirectional cameras," in *IEEE International Conference on Intelligent Robots and Systems (IROS 2006)*, Oct. 2007.

[70] D. Seo, H. Park, K. Jo, K. Eom, S. Yang, and T. Kim, "Omnidirectional stereo vision based vehicle detection and distance measurement for driver assistance system," in *IEEE Industrial Electronics Society*, Vienna, 2013.

[71] P. H. Dwyer and G. R. Southam, "Dual camera mount for stereo imaging," *U.S. Patent,* vol. 6701081, no. B1, Jun. 2000.

[72] W. Benrhaiem, A. S. Hafid, and P. K. Sahu., "Multi-Hop reliability for broadcast-based VANET in city environments," in *2016 IEEE International Conference on Communications (ICC)*, Kuala Lumpur, Malaysia, 2016.

[73] M. A. Togou, A. Hafid, and L. Khoukhi., "SCRP: Stable CDS-based Rrouting protocol for urban vehicular Ad Hoc networks," *IEEE Transactions on Intelligent Transportation Systems,* vol. 17, no. 5, pp. 1298-1307, 2016.

[74] F. Cun-Qun, W. Shang-guang, G. W. Zhe, S. Qi-bo, and Y. Fang-Chun, "Enhanced-throughput multipath routing algorithm based on network coding in IoVs," *Journal on Communications,* vol. 34, no. Z1, pp. 133-141, 2013.

[75] F. Cun-qun, W. Shang-guang, S. Qi-bo, W. Hong-man, Z. Guang-wei, and Y. Fang-chun, "A trust evaluation method of sensor based on energy monitoring," *ACTA Electronica Sinica,* vol. 41, no. 4, pp. 646-651, 2013.

[76] "National strategy for trusted identities in cyberspace (NSTIC): Enhancing online choice, efficiency, security, and privacy," The White House, US., 2011.

[77] "Installation of GPS in buses and autos, document ODR(2010)/75/8, transport department," GOVT of NCT of Delhi, 2010.

[78] European Commission, "Digital signal market strategy," 2015. [Online]. Available: http://ec.europa.eu/priorities/digital-single-market/.

[79] Google, "Open Automobile Alliance," 2015. [Online]. Available: http://www.openautoalliance.net/.

[80] Apple, "CarPlay," 2014. [Online]. Available: http://www.apple.com/.

[81] N. T. Le, M. A. Hossain, A. Islam, D. Y. Kim, Y. J. Choi, and Y. M. Jang, "Survey of promising technologies for 5G networks," *Mobile Information Systems,* vol. 2016, no. 2676589, pp. 1-25, 2016.

[82] S. Ren, K. He, R. Girshick, and J. Sun, "Faster R-CNN: Towards real-time object detection with region proposal networks," *IEEE Transactions on Pattern Analysis and Machine Intelligence,*

vol. 39, no. 6, pp. 1137-1149, Jun. 2017.

[83] B. A. A. Nunes, M. Mendonca, X. N. Nguyen, K. Obraczka, and T. Turletti, "A Survey of Software-Defined Networking: Past, Present, and Future of Programmable Networks," *IEEE Communications Surveys & Tutorials,* vol. 16, no. 3, pp. 1617-1634, 2014.

[84] N. McKeown et al., "Openflow: enabling innovation in campus networks," *ACM SIGCOMM Computer Communication Review,* vol. 38, no. 2, pp. 69-74, Apr. 2008.

[85] L. I. B. López, Á L. V. Caraguay, L. J. G. Villalba, and D. López, "Trends on Virtualisation with software defined networking and network function Virtualisation," *IET Networks,* vol. 4, no. 5, pp. 255-263, Sep. 2015.

[86] N. Feamster, J. Rexford, and E. Zegura, "The road to SDN: an intellectual history of programmable networks," *ACM SIGCOMM Computer Communication Review,* vol. 11, no. 12, Dec. 2013.

[87] A. Lara, A. Kolasani, and B. Ramamurthy, "Network innovation using openflow: A survey," *IEEE Communications Surveys & Tutorials,* vol. 16, no. 1, p. 493–512, 2013.

# List of Publications

❖ **Journal Papers:**

1. **Amirul Islam**, M. T. Hossan, and Y. M. Jang, "Near-Infrared-Based Optical Camera Communication System for Intelligent Internet of Vehicles," *International Journal of Distributed Sensor Networks* (Submitted Feb. 2017).

2. **Amirul Islam**, M. T. Hossan, T. Nguyen, and Y. M. Jang, "Adaptive Spatial-Temporal Resolution Optical Vehicular Communication System using Image Sensor," *International Journal of Distributed Sensor Networks* (Submitted May 2017).

3. M. Z. Chowdhury, M. T. Hossan, **Amirul Islam**, and Y. M. Jang, "Survey of Optical Wireless Technologies: Architectures and Applications," *IEEE Access*, May 2017. (Submitted).

4. M. T. Hossan, **Amirul Islam**, and Y. M. Jang, "Design of an Intelligent Universal Driver Circuit for LED Lights," *The Journal of Korean Institute of Communications and Information Sciences*. (Submitted may 2017).

5. T. Nguyen, **Amirul Islam**, T. Yamazato, and Y. M. Jang, "Technical Issues on IEEE 802.15.7m Task Group and Image Sensor Communication Technologies," *IEEE Communications Magazine* (Accepted).

6. T. Nguyen, **Amirul Islam**, M. T. Hossan, and Y. M. Jang, "Current Status and Performance Analysis of Optical Camera Communication Technologies for 5G Networks," *IEEE Access*, vol. 5, pp. 4574 – 4594, 2017.

7. T. Nguyen, **Amirul Islam**, and Y. M. Jang, "Region-of-Interest Signaling Vehicular System using Optical Camera Communications," *IEEE Photonics Journal*, vol. 9, no. 1, pp. 1-20, Feb. 2017.

8. N. T. Le, M. A. Hossain, **Amirul Islam**, D.Y. Kim, Y. J. Choi, and Y. M. Jang, "Survey of Promising Technologies for 5G Networks," *Mobile Information Systems*, vol. 2016, Article ID 2676589, 25 pages, 2016.

9. M. A. Hossain, **Amirul Islam**, N. T. Le, H. W. Lee, Y. T. Lee, and Y. M. Jang, "Performance Analysis of Smart Digital Signage System Based on software-defined IoT and Invisible Image Sensor Communication," *International Journal of Distributed Sensor Networks*, vol. 12, no. 7, pp. 1–14, Jul. 2016.

10. **Amirul Islam**, M. A. Hossain, N. T. Le, C. H. Hong, and Y. M. Jang, "Robust Software-Defined Scheme for Image Sensor Network," *The Journal of Korean Institute of Communications and Information Sciences*, vol. 41, no. 2, pp. 215-221, Feb. 2016.

11. M. A. Hossain, N. T. Le, **Amirul Islam**, C. H. Hong, and Y. M. Jang, "Invisible Watermarking Based Optical Wireless Communications," *The Journal of Korean Institute of Communications and Information Sciences*, vol. 41, no. 2, pp. 198-205, Feb. 2016.

12. T. Nguyen, C. H. Hong, **Amirul Islam**, N. T. Le, and Y. M. Jang, "Flicker-Free Spatial-PSK Modulation for Vehicular ISC System based on Neural Network," *The Journal of Korean Institute of Communications and Information Sciences*, vol. 41, no. 8, pp. 843-850, Aug. 2016.

❖ **Conference Papers:**

1. **Amirul Islam**, M. T. Hossan, and Y. M. Jang, "Introduction of Optical Camera Communication for Internet of Vehicles (IoV)," to be appeared on 2017 *Ninth International Conference on Ubiquitous and Future Networks (ICUFN)*. (Accepted).

2. M. T. Hossan, M. Z. Chowdhury, **Amirul Islam**, and Y. M. Jang, "Simplified Photogrammetry Using Optical Camera Communication for Indoor Positioning," to be appeared on 2017 *Ninth International Conference on Ubiquitous and Future Networks (ICUFN)*. (Accepted).

3. **Amirul Islam**, M. T. Hossan, M. Z. Chowdhury, and Y. M. Jang, "An Indoor Localization Scheme Based on Integrated Artificial Neural Fuzzy Logic," The 14th *IEEE Vehicular Technology Society Asia Pacific Wireless Communications Symposium (APWCS)*, Incheon, Korea, Aug. 2017 (Submitted Apr. 2017).

4. **Amirul Islam**, M. A. Hossain, T. Nguyen, and Y. M. Jang, "High Temporal-Spatial Resolution Optical Wireless Communication Technique using Image Sensor," in Proc. of 2016 *International Conference on Information and Communication Technology Convergence (ICTC)*, Jeju Island, South Korea, 2016, pp. 1165-1169.

5. M. A. Hossain, **Amirul Islam**, and Y. M. Jang, "Effects of viewing angle between camera and display in invisible image sensor communication," in Proc. of 2016 *International Conference on Information and Communication Technology*

*Convergence (ICTC)*, Jeju Island, South Korea, 2016, pp. 1221-1223.

6. **Amirul Islam**, M. A. Hossain, and Y. M. Jang, "Interference Mitigation Technique for Time-of-Flight (ToF) Camera," in Proc. of 2016 *Eighth International Conference on Ubiquitous and Future Networks (ICUFN)*, Vienna, pp. 134-139, 2016.

7. M. A. Hossain, **Amirul Islam**, and Y. M. Jang, "Pixel to Signal Conversion Based Invisible Image Sensor Communication," in Proc. of 2016 *Eighth International Conference on Ubiquitous and Future Networks (ICUFN)*, Vienna, pp. 448-452, 2016.

8. T. Nguyen, N. T. Lee, M. A. Hossain, C. H. Hong, **Amirul Islam**, and Y. M. Jang, "Flicker-Free Spatial-PSK Modulation Scheme for Vehicular Image Sensor Communications," in Proc. of 2016 *Eighth International Conference on Ubiquitous and Future Networks (ICUFN)*, Vienna, pp. 128-133, 2016.

9. **Amirul Islam**, M. A. Hossain, N. T. Le, Y. T. Lee, H. W. Lee, and Y. M. Jang, "Software Defined Networking Enabled Digital Signage for Disaster Management," in Proc. of *2015 KICS Korea and Southeast Asia ICT International Workshop*, Siem Reap, Cambodia, 21~23 Dec. 2015.

10. **Amirul Islam**, M. T. Hossan, E. Shin, C. S. Kim, and Y. M. Jang, "Optical Camera Communication and Color Code based Digital Signage Service Implementation," in Proc. *General Conference of KICS* (*Summer*), Jun. 2017, Jeju Island, Korea.

11. M. T. Hossan, **Amirul Islam**, V. T. Luan and Y. M. Jang, "IEEE 802.15.7m Standardization: Current Status and Application," in Proc. of the 27th *Joint Conference on Communications and Information (JCCI)*, Apr. 2017.

12. M. Z. Chowdhury, M. T. Hossan, **Amirul Islam**, and Y. M. Jang, "Coexistence of RF and VLC Systems for 5G and Beyond Wireless Communications," in Proc. of the 27th *Joint Conference on Communications and Information (JCCI)*, Apr. 2017.

13. **Amirul Islam**, M. T. Hossan, T. L. Vu, and Y. M. Jang, "Opportunities and Scopes in IEEE 802.15 Vehicular Assistant Technology Interest Group," in Proc. of *General Conference of KICS* (W*inter*), Jan. 2017, pp. 1295-1296.

14. M. T. Hossan, **Amirul Islam**, N. V. Phuoc, and Y. M. Jang, "Precious Indoor Localization using Optical Camera Communication for Smartphone," in Proc. of *General Conference of KICS* (*Winter*), Jan. 2017, pp. 1294-1295.

15. **Amirul Islam**, M. A. Hossain, N. T. Le, C. H. Hong, T. Nguyen, and Y. M. Jang, "Precise Localization in Tunnel Environment using Visible Light Communication," in Proc. of the 26th *Joint Conference on Communications and Information (JCCI)*, Apr. 2016.

16. **Amirul Islam**, M. A. Hossain, C. H. Hong, and Y. M. Jang, "Image Sensor based Indoor Localization Scheme using Fuzzy Logic in Ambient Intelligent Environments," in Proc. of the 26th *Joint Conference on Communications and Information (JCCI)*, Apr. 2016.

17. M. A. Hossain, **Amirul Islam**, N. T. Le, T. Nguyen, C. H. Hong, and Y. M. Jang, "Opportunities from Invisible Watermarking for Device-to-Device Communication," in Proc. of the 26th *Joint Conference on Communications and Information (JCCI)*, Apr. 2016.

18. M. A. Hossain, **Amirul Islam**, C. H. Hong, N. T. Le, T. Nguyen, and Y. M. Jang, "Similarity Analysis for Invisibility in Image Sensor Communication," in Proc. of the 26th *Joint Conference on Communications and Information (JCCI)*, Apr. 2016.

19. N. T. Le, M. A. Hossain, **Amirul Islam**, C. H. Hong, T. Nguyen, and Y. M. Jang, "Consideration Issues for MIMO Camera Communications," in Proc. of the 26th *Joint Conference on Communications and Information* (*JCCI*), Apr. 2016.

20. T. Nguyen, N. T. Le, M. A. Hossain, C. H. Hong, **Amirul Islam**, and Y. M. Jang, "A Review of Image Sensor Communications Recent Activities in IEEE 802.15.7r1 Task Group (TG7r1)," in Proc. of the 26th *Joint Conference on Communications and Information* (*JCCI*), Apr. 2016.

21. **Amirul Islam**, M. A. Hossain, and Y. M. Jang, "SVM-based Positioning Approach in Image Sensor Network," in Proc. of *General Conference of KICS* (*Winter*), Jan. 2016.

22. C. H. Hong, M. A. Hossain, Nam Tuan Le, Trang Nguyen, **Amirul Islam**, and Y. M. Jang, "Alpha Channel based Invisible Image Sensor Communication," in Proc. of *General Conference of KICS* (*Winter*), pp. 777-778, Jan. 2016.

23. M. A. Hossain, **Amirul Islam**, N. T. Le, C. H. Hong, T. Nguyen, and Y. M. Jang, "Detection Issue for Moving Drone in Wireless Sensor Networks," in Proc. of *General Conference of KICS* (*Fall*), pp. 214-215, 2015.

❖ **Contribution to IEEE Standardization**

1. **Amirul Islam**, M. T. Hossan, and Y. M. Jang, "Interference Characterization of Artificial Light Sources for long range OCC," 07 May 2017. [Available] Online:

https://mentor.ieee.org/802.15/dcn/17/15-17-0276-00-0vat-interference-characterizationof-artificial-light-sources-for-long-range-occ.pptx

2. M. T. Hossan, **Amirul Islam**, and Y. M. Jang, "IEEE 802.15 IG VAT: Using rear light for long range OCC to ensure safety issues," 07 May 2017.[Available] Online: https://mentor.ieee.org/802.15/dcn/17/15-17-0275-00-0vat-ieee-802-15-ig-vat-using-rear-light-for-long-range-occ-to-ensure-safety-issues.pptx

3. **Amirul Islam**, Y. M. Jang, "Use Cases and Functional Requirements for IEEE 802.15 Vehicular Assistant Technology (VAT) IG," 18 Jan. 2017. [Available] Online: https://mentor.ieee.org/802.15/dcn/17/15-17-0061-00-0vat-use-cases-and-functional-requirements-for-ieee-802-15-vehicular-assistant-technology-vat-ig.pptx

4. **Amirul Islam,** M. A. Hossain, and Y. M. Jang, "Kookmin University Response to Call for Interests (CFI) of IEEE 802 IG-DEP," 10 Nov. 2015. [Available] Online: https://mentor.ieee.org/802.15/dcn/15/15-15-0895-00-0dep-kookmin-university-response-to-call-for-interests-cfi-of-ieee-802-ig-dep.pptx

5. Y. M. Jang, **Amirul Islam**, "Opening Long Range Automotive OWC Interest Group," 9 Nov. 2016. [Available] Online: https://mentor.ieee.org/802.15/dcn/16/15-16-0793-01-wng0-opening-long-range-automotive-owc-interest-group.pdf

6. Y. M. Jang, **Amirul Islam**, "Opening Long Range Automotive OWC Interest Group," 8 Nov. 2016. [Available] Online: https://mentor.ieee.org/802.15/dcn/16/15-16-0793-00-wng0-opening-long-range-automotive-owc-interest-group.pdf

7. Y. M. Jang, T. Nguyen, N. T. Le, M. S. Ifthekhar, M. A. Hossain, C. H. Hyun, and **Amirul Islam**, "General Architecture of Kookmin University PHY and MAC multiple-proposal for Image Sensor Communication," 10 Jan. 2016. [Available] Online: https://mentor.ieee.org/802.15/dcn/16/15-16-0011-00-007a-general-architecture-of-kookmin-university-phy-and-mac-multiple-proposal-for-image-sensor-communication.pptx

8. Y. M. Jang, T. Nguyen, N. T. Le, M. S. Ifthekhar, M. A. Hossain, C. H. Hyun, and **Amirul Islam**, "General Architecture of Kookmin University PHY and MAC multiple-proposal for Image Sensor Communication," 18 Jan. 2016. [Available] Online: https://mentor.ieee.org/802.15/dcn/16/15-16-0011-01-007a-general-architecture-of-kookmin-university-phy-and-mac-multiple-proposal-for-image-sensor-communication.pptx